\documentclass[11pt]{article}

\usepackage[margin=1in]{geometry}
\usepackage{amsmath,amssymb,amsfonts}
\usepackage{booktabs}
\usepackage{array}
\usepackage{enumitem}
\usepackage{graphicx}
\usepackage{placeins}
\usepackage{xcolor}
\usepackage[round,authoryear]{natbib}
\setcitestyle{aysep={}}
\usepackage{hyperref}
\usepackage{setspace}
\usepackage{authblk}
\graphicspath{{figures/}}

\hypersetup{
  colorlinks=true,
  linkcolor=blue,
  citecolor=blue,
  urlcolor=blue
}

\newcommand{\E}{\mathbb{E}}
\newcommand{\ii}{\mathrm{i}}

\title{COS--TT--CHF: A Tensor-Train Characteristic-Function COS Method for Multi-Asset Option Pricing}
\author[1]{Lucas Arenstein\thanks{Email: \href{mailto:lsa@di.ku.dk}{lsa@di.ku.dk}}}
\author[1]{Michael Kastoryano}
\affil[1]{Department of Computer Science, University of Copenhagen, Denmark}
\date{\today}

\begin{document}

\maketitle

\begin{abstract}
This paper considers European multi-asset option pricing under
Lévy and affine characteristic-function models. The main obstruction is the curse of
dimensionality: direct multidimensional COS pricing forms tensor-product
coefficient arrays whose size grows exponentially with the number of assets. We
study and extend
COS--TT--CHF, a low-rank construction that uses TT-cross to compress sampled
characteristic-function tensors into tensor-train COS coefficients for
arithmetic basket and min/max option pricing. Once built, the compressed
representation gives fast post-setup strike-grid and selected component
Delta/Vega calculations. The numerical study compares with
adaptive-quadrature Fourier benchmarks, direct COS,
a tensor-Fourier min-option benchmark, and quasi-Monte Carlo (QMC) references
based on randomized Sobol points. The
reported timings show a low-dimensional crossover against direct COS as the
benchmark moves from \(d=2\) to \(d=4\), favorable timings against
the tensor-Fourier min-option benchmark from \(d=3\) onward, and favorable
timings against the QMC common-Heston reference already at \(d=2\). The reported tests
reach \(d=30\) for GBM and \(d=20\) for VG, NIG, and common-Heston benchmark
families, with accuracy, rank, runtime, control-sensitivity, and component Delta/Vega
diagnostics reported throughout.
\end{abstract}

\noindent\textbf{Keywords:}
Multi-asset option pricing, Fourier-COS methods, characteristic functions,
tensor trains, low-rank tensor approximation, TT-cross approximation, basket
options, min/max options, Greeks.

\section{Introduction}
European option pricing under characteristic-function models is a natural
setting for Fourier methods, including transform-based FFT pricing
\citep{carrmadan1999}. In one dimension, the COS method gives an exceptionally
accurate and efficient route from the characteristic function to cosine
coefficients on a truncated interval \citep{fang2008}. Related Fourier/COS developments include
SINC pricing for one-dimensional characteristic-function models
\citep{baschetti2022}, COS methods for regime-switching time-changed Lévy
models \citep{tour2018}, and the damped COS method for multivariate
distribution functions and European option prices \citep{junike2025}.

The difficulty in multi-asset pricing is dimensional.  Direct multidimensional
Fourier or COS constructions represent densities, coefficients, or pricing
integrands on tensor-product grids. With \(n\) points per dimension, a full
tensor has \(n^d\) entries, so these constructions become expensive as the
number of assets grows.  Recent
multivariate Fourier/COS and adaptive quadrature methods address this issue in
different ways: payoff smoothing can improve adaptive sparse-grid integration
for basket options \citep{bayer2018}, optimized damping and hierarchical
adaptive quadrature target multi-asset Fourier pricing in Lévy models
\citep{bayer2023}, quasi-Monte Carlo (QMC), in randomized form, can replace tensor-product
quadrature after a Fourier-domain transformation \citep{bayer2024rqmc}, and
damped COS formulas provide another multivariate characteristic-function route
\citep{junike2025}.

These QMC and adaptive-quadrature methods are important comparators, but they
address a different numerical object. Sampling methods estimate expectations
under a chosen sampling protocol, while COS methods construct spectral
coefficient representations from the characteristic function. The motivation
for COS--TT--CHF is to make this representation usable in higher dimensions: if
the sampled characteristic-function tensor admits a moderate-rank TT
approximation, then the resulting COS coefficient TT can be reused for strike
grids, payoff-specific contractions, diagnostics, and selected first-order
Greeks. In this sense, COS--TT--CHF is not only a coefficient-compression
device but a characteristic-function-based solver whose reusable state is a
compressed COS representation.

Tensor-network methods provide another way to attack the same tensor-product
growth by approximating high-dimensional arrays in low-rank format.  In option
pricing, this idea has been used for tensor-Fourier integrands and related
multi-asset pricing problems \citep{kastoryano2022}, for learning parameter
dependence in Fourier-based multi-asset pricing \citep{sakurai2025parameters}, and for
full-grid Black--Scholes PDE solvers based on quantics tensor trains
\citep{arenstein2026fullgrid}.  This paper studies COS--TT--CHF, a characteristic-function-side tensor-train
variant of the COS method. Schaap's COS--TT first builds or samples the
multidimensional Fourier--cosine coefficient tensor and compresses that tensor
in TT format \citep{schaapthesis}. COS--TT--CHF instead moves the low-rank
approximation upstream by compressing the sampled joint characteristic function
and then transforming the resulting TT cores into COS coefficient cores.

This shift changes the main numerical bottleneck. Rather than approximating the
coefficient tensor directly, one must resolve a smooth but oscillatory
frequency-domain object. Schaap reported that inadequate frequency resolution
can prevent an accurate TT approximation of this characteristic-function
tensor.

We revisit that construction and make the following contributions:
\begin{itemize}[leftmargin=*]
  \item We give a self-contained COS--TT--CHF workflow: sample the joint
  characteristic function on a frequency tensor grid, compress the sampled
  tensor in TT format, and convert the compressed object corewise into joint COS
  coefficients.

  \item We extend the payoff treatment beyond arithmetic baskets. Basket
  prices are obtained through a one-dimensional basket projection, while
  min/max prices are obtained from rectangle probabilities and tail integrals
  computed directly from the joint COS coefficient TT.

  \item We evaluate the method on basket, min/max, multi-strike, and selected
  Greek calculations, reporting pricing errors, ranks, characteristic-function
  evaluations, runtimes, and control-sensitivity diagnostics.
\end{itemize}

The main numerical findings are as follows. The reported experiments reproduce
published basket benchmarks and the tensor-Fourier min-option benchmark
\citep{kastoryano2022}
within the stated validation tolerances. The high-dimensional tests reach
\(d=30\) for GBM and \(d=20\) for VG, NIG, and common-Heston benchmark
families. The numerical results also show a low-dimensional runtime crossover.
In the local direct-COS comparison, COS--TT--CHF is slower at \(d=2\) but
faster at \(d=4\). In the min-option bridge and common-Heston stress test, the
reported timings favor COS--TT--CHF in the tested dimensions beyond \(d=2\),
with the comparison conventions stated in the corresponding tables. Once the
compressed representation is built, additional strikes and the reported
component Delta/Vega calculations are fast post-setup computations that avoid a
new TT-cross construction.

The rest of the paper is organized as follows. Section~2 fixes the pricing
setup and payoff classes. Sections~3 and~4 describe the COS--TT--CHF
construction and its conversion into basket and min/max prices. Section~5
treats selected Greek computations, and Section~6 discusses numerical error,
diagnostics, and cost. The numerical sections then give the benchmark protocol,
validation results, high-dimensional experiments, multi-strike reuse, and Greek
validation.

\section{Pricing Setup, Payoff Suite, and Model Class}
All prices are computed under a risk-neutral probability measure, under which
the discounted asset prices \(e^{-rt}S_i(t)\) are martingales.  The terminal
asset vector is \(S_T=(S_1(T),\ldots,S_d(T))\), and the log-price vector is
\[
  X_T=(X_1,\ldots,X_d)
  =
  (\log S_1(T),\ldots,\log S_d(T)).
\]
The method assumes that the joint characteristic function
\begin{equation*}
  \varphi_X(\omega)
  =
  \E\!\left[\exp(\ii\omega^\top X_T)\right],
  \qquad \omega\in\mathbb{R}^d,
\end{equation*}
can be evaluated pointwise.  This assumption is the common input to all
COS--TT--CHF calculations below.

When \(X_T\) admits a density, denote it by \(f_X\).  Under the usual Fourier
inversion conditions,
\begin{equation}
  f_X(x)
  =
  \frac{1}{(2\pi)^d}
  \int_{\mathbb{R}^d}
  \varphi_X(\omega)e^{-\ii\omega^\top x}\,d\omega,
  \qquad x\in\mathbb{R}^d.
  \label{eq:fourier-inversion}
\end{equation}
The COS method does not evaluate this integral directly.  Instead, it uses
\(\varphi_X\) to compute cosine coefficients of \(f_X\) on a finite log-price
box.  COS--TT--CHF applies the same idea before the full multidimensional COS
coefficient tensor is formed.

Define the arithmetic basket map
\[
  \mathcal{H}(x)=\sum_{j=1}^d w_j e^{x_j},
\]
where \(w_j\) are fixed basket weights. Write the terminal basket value as
\(H=\mathcal{H}(X_T)=\sum_{j=1}^d w_j S_j(T)\).
For strike \(K\), the discounted basket call and put prices are
\begin{align*}
  C(K)&=e^{-rT}\E[(H-K)^+],
  &
  P(K)&=e^{-rT}\E[(K-H)^+].
\end{align*}

The scalar basket variable \(H\) separates the payoff integration from the
high-dimensional density approximation. We first write the price as a
one-dimensional integral in the basket variable \(h\). Only then do we explain
how the COS expansion of the high-dimensional density \(f_X\) is used to
compute the basket characteristic function \(\varphi_H\).

\subsection{Characteristic-Function Model Class}
Here ``model class'' refers to the risk-neutral law of the terminal log-price
vector \(X_T\), represented by the joint characteristic function
\(\varphi_X\) used by the pricing method.  The method does not depend on a
particular asset model beyond access to \(\varphi_X\) on the frequency grid.
The numerical study uses four characteristic-function families to test
different regimes: multivariate geometric Brownian motion (GBM), variance
gamma (VG), normal inverse Gaussian (NIG), and a common-factor affine Heston
model.  GBM provides the Gaussian baseline.  VG and NIG test non-Gaussian
terminal laws with jumps or heavy tails.  The common-Heston case tests a
stochastic-volatility setting where the terminal characteristic function is
available but path-based QMC must also assess time-step sensitivity.

For GBM, \(X_T\) is Gaussian with the usual risk-neutral mean and covariance,
so the joint characteristic function is available in closed form.  For VG, we
use the standard time-changed Brownian specification with the drift correction
chosen so that the discounted assets are martingales.  The generated GBM and
VG parameter rules, the NIG and Heston conventions, and all source-paper
validation conventions are stated in the benchmark-protocol tables and
appendices.  We keep these conventions separate because the OD+ASGQ,
Junike--Stier, Kastoryano--Pancotti, and generated experiments do not use the
same payoff, weighting, rate, or dependence conventions.

Thus the method section below is written for a generic evaluable
characteristic function.  Model-specific formulas are used only to generate
the numerical entries of \(\varphi_X\). They are not part of the tensor-train
compression argument.
The construction below therefore treats the characteristic function as the model
input and the compressed COS coefficient tensor as the reusable computational
object. This separation is useful because the same compressed tensor can then be
reused for basket projection, min/max payoff contractions, diagnostics, and
selected derivative calculations.

\subsection{Reduction to the Scalar Basket Variable}
The random scalar terminal basket value is
\(
  H := \mathcal{H}(X_T).
\)
When the law of \(H\) admits a density \(f_H\), the call price satisfies
\begin{align}
  C(K)
  &=
  e^{-rT}\E[(H-K)^+]                                                   \notag\\
  &=
  e^{-rT}\int_{\mathbb{R}} (h-K)^+ f_H(h)\,dh .
  \label{eq:basket-price-density}
\end{align}
This is the clean target: once the one-dimensional basket density \(f_H\) is
available, basket pricing is a standard one-dimensional payoff integration
problem.

\paragraph{Push-forward view.}
The density \(f_H\) is the push-forward of \(f_X\) under the
map \(\mathcal{H}:\mathbb{R}^d\to\mathbb{R}\).  In formal distributional notation,
\begin{equation*}
  f_H(h)
  =
  \int_{\mathbb{R}^d}
  \delta\bigl(h-\mathcal{H}(x)\bigr) f_X(x)\,dx .
\end{equation*}
Equivalently, for regular values \(h\) and under the usual regularity
assumptions for the coarea formula,
\begin{equation}
  f_H(h)
  =
  \int_{\mathcal{H}^{-1}(h)}
  \frac{f_X(x)}{\|\nabla \mathcal{H}(x)\|}\,dS(x).
  \label{eq:basket-coarea-density}
\end{equation}
This identity records the geometric factor in the change from the
high-dimensional variable \(x\) to the scalar basket value \(h\): the level set
\(\mathcal{H}^{-1}(h)\) carries the surface measure \(dS\), and the Jacobian
factor \(\|\nabla \mathcal{H}\|^{-1}\) appears.  Since evaluating
\eqref{eq:basket-coarea-density} directly is not the numerical route used
here, we instead compute the characteristic function of \(H\), and then
reconstruct \(f_H\) by a
one-dimensional COS expansion.

\subsection{Standard One-Dimensional COS Pricing Once \texorpdfstring{\(\varphi_H\)}{phiH} is Known}
Let \([a_H,b_H]\) be the one-dimensional COS truncation interval for the
basket value \(H\). The implementation uses a cumulant-based interval for
\(H\), enlarged when necessary to contain the strike grid. The resulting
reconstruction is checked through density mass, first moment, and put--call
parity diagnostics. See Section~\ref{sec:pricing-from-cos-coef}.
Define the standard COS frequencies
\begin{equation*}
  u_n=\frac{n\pi}{b_H-a_H},
  \qquad n=0,\ldots,N_H-1 .
\end{equation*}
If the basket characteristic function
\(
  \varphi_H(u)=\E\!\left[e^{\ii uH}\right]
\)
is known, the COS coefficients of \(f_H\) on \([a_H,b_H]\) are approximated by
\begin{equation*}
  F_n
  =
  \frac{2}{b_H-a_H}
  \operatorname{Re}\!\left[
    \varphi_H(u_n)e^{-\ii u_n a_H}
  \right].
\end{equation*}
Then
\begin{equation}
  f_H(h)
  \approx
  \sum_{n=0}^{N_H-1}{}'
  F_n
  \cos\!\left(n\pi\frac{h-a_H}{b_H-a_H}\right),
  \label{eq:basket-density-cos-reconstruction}
\end{equation}
where the prime means that the \(n=0\) term in the sum is multiplied by \(1/2\).
For \(K\in[a_H,b_H]\), substitution into \eqref{eq:basket-price-density} gives
\begin{equation*}
  C(K)
  \approx
  e^{-rT}\int_K^{b_H} (h-K) f_H(h)\,dh .
\end{equation*}
For many strikes, it is convenient to precompute the prefix integrals
\[
  L_0(K)=\int_{a_H}^{K} f_H(h)\,dh,
  \qquad
  L_1(K)=\int_{a_H}^{K} h f_H(h)\,dh.
\]
Writing \(T_0=L_0(b_H)\) and \(T_1=L_1(b_H)\), the truncated-interval call
and put prices, for strikes in the basket interval, are
\begin{align}
  C(K)
  &\approx
  e^{-rT}\bigl[(T_1-L_1(K))-K(T_0-L_0(K))\bigr],
  \label{eq:basket-call-prefix}\\
  P(K)
  &\approx
  e^{-rT}\bigl[KL_0(K)-L_1(K)\bigr].
  \label{eq:basket-put-prefix}
\end{align}
Thus \(f_H\) is reconstructed once, and each additional strike
requires only one-dimensional evaluation of these cumulative quantities.  For
strikes outside \([a_H,b_H]\), the implementation uses the corresponding
endpoint values of the cumulative integrals.  It remains to compute
\(\varphi_H\) at the scalar COS frequencies.

\section{From Characteristic Functions to COS Coefficients}
The previous section reduces basket pricing to the evaluation of
\(\varphi_H\) at a one-dimensional COS frequency grid.  This section derives
that quantity from the joint density \(f_X\), inserts the multidimensional COS
expansion, and then explains how the required COS coefficients are obtained
from the joint characteristic function \(\varphi_X\).

\subsection{Computing \texorpdfstring{\(\varphi_H\)}{phiH} from the Joint Density of \texorpdfstring{\(X_T\)}{XT}}
By definition,
\begin{align*}
  \varphi_H(u)
  =
  \E\!\left[e^{\ii uH(X_T)}\right]                                      
  =
  \int_{\mathbb{R}^d}
  e^{\ii u\mathcal{H}(x)} f_X(x)\,dx .
\end{align*}
After truncating the log-price domain to
\(
  D=\prod_{m=1}^d [a_m,b_m],
\)
we use the truncation approximation
\begin{equation}
  \varphi_H(u)
  \approx
  \int_D e^{\ii u\mathcal{H}(x)} f_X(x)\,dx .
  \label{eq:basket-chf-truncated}
\end{equation}
The approximation of this integral uses a tensor-product COS representation of
\(f_X\) on \(D\).  In the reported experiments, the box is chosen coordinatewise from the marginal
mean and variance of each terminal log-price. Let
\(\mu_m=\E[X_m]\) and \(\Sigma_{mm}=\operatorname{Var}(X_m)\), 
and given a truncation scale \(L_{\rm tr}>0\), we set 
\begin{equation}
  a_m=\mu_m-L_{\rm tr}\sqrt{\Sigma_{mm}},
  \qquad
  b_m=\mu_m+L_{\rm tr}\sqrt{\Sigma_{mm}},
  \qquad m=1,\ldots,d.
  \label{eq:log-price-truncation-box}
\end{equation}
The scale \(L_{\rm tr}\) is treated as a numerical resolution parameter and is
checked through the numerical diagnostics reported below.

\subsection{Define the COS Expansion for \texorpdfstring{\(f_X(x)\)}{fX(x)}}
For each coordinate \(m\), define the one-dimensional COS basis
\begin{equation*}
  \psi_k^{(m)}(x)
  =
  \cos\!\left(k\pi\frac{x-a_m}{b_m-a_m}\right),
  \qquad
  k=0,\ldots,N_m-1 .
\end{equation*}
On \(D\), approximate the joint density by the tensor-product COS expansion
\begin{equation}
  f_X(x)
  \approx
  \sum_{k_1=0}^{N_1-1}{}'\cdots
  \sum_{k_d=0}^{N_d-1}{}'
  c_{\mathbf{k}}
  \prod_{m=1}^d \psi_{k_m}^{(m)}(x_m),
  \qquad
  \mathbf{k}=(k_1,\ldots,k_d).
  \label{eq:joint-cos-expansion}
\end{equation}
Throughout the paper, a prime on a one-dimensional COS sum halves the zero mode,
\(
  \sum_{k=0}^{N-1}{}' a_k
  :=
  \frac{a_0}{2}+\sum_{k=1}^{N-1}a_k .
\)
In the tensor-product expansion above, this convention is applied separately
in each coordinate. The coefficient array \(\mathcal{C}=(c_{\mathbf{k}})\) is a
\(d\)-dimensional tensor.
If \(N_m=N\) for all \(m\), a direct representation contains \(N^d\)
coefficients.  This exponential growth is the computational obstruction that
prevents direct multidimensional COS from being used as a high-dimensional
pricing engine.

Substitution of \eqref{eq:joint-cos-expansion} into
\eqref{eq:basket-chf-truncated} gives
\begin{align*}
  \varphi_H(u)
  &\approx
  \sum_{k_1=0}^{N_1-1}{}'\cdots
  \sum_{k_d=0}^{N_d-1}{}'
  c_{\mathbf{k}}
  \int_D
  e^{\ii u\sum_{m=1}^d w_m e^{x_m}}
  \prod_{m=1}^d \psi_{k_m}^{(m)}(x_m)\,dx .
\end{align*}
Because \(\mathcal{H}(x)=\sum_{m=1}^d w_m e^{x_m}\), the exponential factor
separates:
\begin{equation*}
  e^{\ii u\sum_{m=1}^d w_m e^{x_m}}
  =
  \prod_{m=1}^d e^{\ii u w_m e^{x_m}} .
\end{equation*}
Thus the box integral factorizes:
\begin{align*}
  \varphi_H(u)
  &\approx
  \sum_{k_1=0}^{N_1-1}{}'\cdots
  \sum_{k_d=0}^{N_d-1}{}'
  c_{\mathbf{k}}
  \prod_{m=1}^d
  \int_{a_m}^{b_m}
  e^{\ii u w_m e^x}\psi_{k_m}^{(m)}(x)\,dx .
\end{align*}
Define the one-dimensional basket-transform factors
\begin{equation*}
  J_m(k,u)
  =
  \int_{a_m}^{b_m}
  e^{\ii u w_m e^x}\psi_k^{(m)}(x)\,dx .
\end{equation*}
For fixed \(m\), \(k\), and \(u\), this integral is a complex scalar. For
fixed \(u\), the collection \(J_m(\cdot,u)\) is a vector over the COS modes in
coordinate \(m\).
Then
\begin{equation}
  \varphi_H(u)
  \approx
  \sum_{k_1=0}^{N_1-1}{}'\cdots
  \sum_{k_d=0}^{N_d-1}{}'
  c_{\mathbf{k}}
  \prod_{m=1}^d J_m(k_m,u).
  \label{eq:basket-chf-from-cos}
\end{equation}
Equation~\eqref{eq:basket-chf-from-cos} expresses \(\varphi_H\) as a
contraction of the COS coefficient tensor with one-dimensional
basket-transform factors.

\subsection{Recovering COS Coefficients from the Joint Characteristic Function}
The coefficient tensor \(\mathcal{C}\) can be obtained without ever evaluating
the density \(f_X\) directly by applying the Fourier Inversion defined in
\eqref{eq:fourier-inversion}.  The COS projection coefficient is
\begin{equation}
  c_{\mathbf{k}}
  =
  \left(\prod_{m=1}^d\frac{2}{b_m-a_m}\right)
  \int_D
  f_X(x)
  \prod_{m=1}^d\psi_{k_m}^{(m)}(x_m)\,dx .
  \label{eq:cos-projection}
\end{equation}
Substituting \eqref{eq:fourier-inversion} into \eqref{eq:cos-projection} and
separating the product box gives
\begin{equation*}
  c_{\mathbf{k}}
  =
  \left(\prod_{m=1}^d\frac{2}{b_m-a_m}\right)
  \frac{1}{(2\pi)^d}
  \int_{\mathbb{R}^d}
  \varphi_X(\omega)
  \prod_{m=1}^d I_m(\omega_m,k_m)\,d\omega,
\end{equation*}
where
\begin{equation*}
  I_m(\omega,k)
  =
  \int_{a_m}^{b_m}
  e^{-\ii\omega x}\psi_k^{(m)}(x)\,dx .
\end{equation*}
The multidimensional dependence remains in \(\varphi_X(\omega)\), while the
transform from frequency variables to COS modes is applied separately in each
coordinate. We therefore approximate the remaining frequency integral by a
product quadrature rule: the characteristic function is sampled on a
tensor-product frequency grid, and the one-dimensional transform factors are
absorbed coordinatewise.

Let \(\omega_{m,j}\) and \(q_{m,j}\), \(j=1,\ldots,M_m\), denote the quadrature
nodes and weights in coordinate \(m\). The one-dimensional frequency rule may
be an ordinary Gauss--Legendre rule on the truncated window or a mapped rule
that concentrates nodes near a chosen frequency center. Write
\begin{equation*}
  \Phi[j_1,\ldots,j_d]
  =
  \varphi_X(\omega_{1,j_1},\ldots,\omega_{d,j_d}) .
\end{equation*}
With the normalization used in the projection formula above, define
\begin{equation}
  \mathcal{M}_m[j,k]
  =
  \frac{q_{m,j}}{\pi(b_m-a_m)}
  I_m(\omega_{m,j},k).
  \label{eq:frequency-to-cos-map}
\end{equation}
Then the discrete coefficient formula is
\begin{equation*}
  c_{k_1,\ldots,k_d}
  \approx
  \sum_{j_1=1}^{M_1}\cdots\sum_{j_d=1}^{M_d}
  \Phi[j_1,\ldots,j_d]
  \prod_{m=1}^d \mathcal{M}_m[j_m,k_m].
\end{equation*}
Together with the COS order \(N_m\), the main discretization controls are the
frequency half-widths \(W_m\) and the quadrature sizes \(M_m\). The frequency grid is treated as a numerical control rather
than a fixed background choice.  The full sampled tensor has
\(\prod_{m=1}^dM_m\) entries, so COS--TT--CHF treats
\(\Phi[j_1,\ldots,j_d]\) as an entry oracle: a requested multi-index is mapped
to one direct characteristic-function evaluation.

\subsection{TT Compression and Contraction}
The next task is to build a TT representation of the sampled
characteristic-function tensor \(\Phi\).  We use the standard tensor-train
format \citep{oseledets2011} and TT-cross
\citep{oseledets2010ttcross}, which constructs a tensor-train approximation
from selected tensor entries.  First, we fix the notation. For a tensor \(A\in\mathbb{F}^{n_1\times\cdots\times n_d}\), a TT
representation has the form
\[
  A[i_1,\ldots,i_d]
  =
  G_1(i_1)G_2(i_2)\cdots G_d(i_d),
\]
where \(G_m(i_m)\in\mathbb{F}^{r_{m-1}\times r_m}\) and \(r_0=r_d=1\).
In COS--TT--CHF, the tensor to be compressed is \(A=\Phi\).  The compressed
sampled CHF tensor is represented as
\begin{equation}
  \widehat{\Phi}[j_1,\ldots,j_d]
  =
  G_1(j_1)G_2(j_2)\cdots G_d(j_d),
  \qquad
  G_m(j_m)\in\mathbb{C}^{r_{m-1}\times r_m}.
  \label{eq:tt-chf}
\end{equation}
Every sampled entry used by the algorithm is still an exact evaluation of
\(\varphi_X\).  After compression, held-out tensor entries are evaluated
directly from \(\varphi_X\) and compared with \(\widehat{\Phi}\).

Apply the one-dimensional maps \(\mathcal{M}_m\) to the TT cores:
\begin{equation}
  B_m(k)
  =
  \sum_{j=1}^{M_m}
  G_m(j)\mathcal{M}_m[j,k],
  \qquad k=0,\ldots,N_m-1.
  \label{eq:cos-coeff-core-transform}
\end{equation}
The scalar map values \(\mathcal{M}_m[j,k]\) change the physical index from
frequency nodes \(j\) to COS modes \(k\), while the TT-rank dimensions of the
core slice are unchanged.
Then
\begin{equation}
  c_{k_1,\ldots,k_d}
  \approx
  B_1(k_1)B_2(k_2)\cdots B_d(k_d).
  \label{eq:cos-coeff-tt}
\end{equation}
The products in \eqref{eq:cos-coeff-tt} are ordered matrix-chain products, and
the boundary ranks \(r_0=r_d=1\) make the result scalar for each
multi-index \((k_1,\ldots,k_d)\).

\begin{figure}[t]
  \centering
  \includegraphics[width=\textwidth]{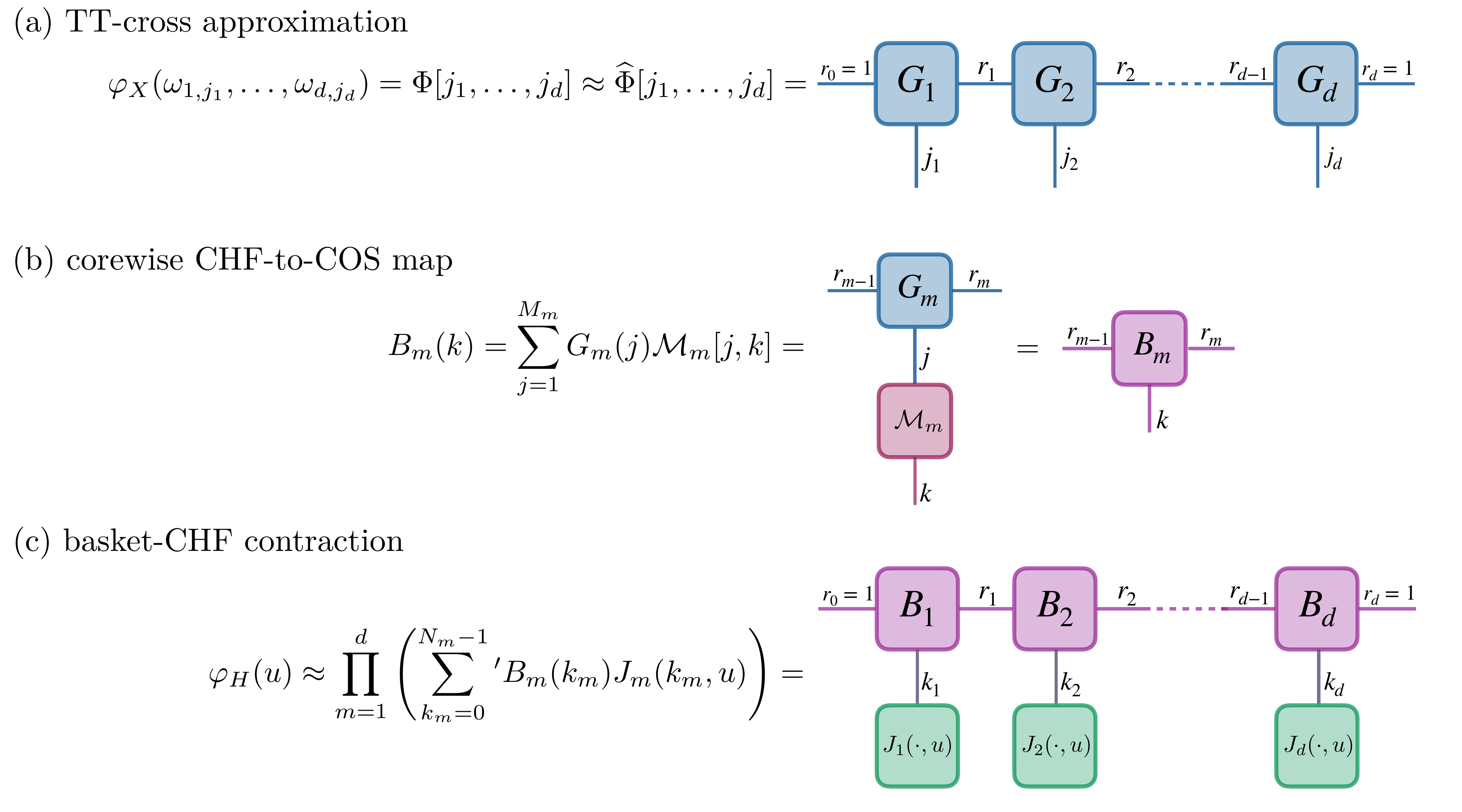}
  \caption{Tensor-network representation of the COS--TT--CHF construction.
  (a) The joint characteristic function sampled on the tensor-product frequency
  grid defines the tensor \(\Phi\), which is approximated by TT-cross as
  \(\widehat{\Phi}\) with TT cores \(G_m\). (b) The one-dimensional
  CHF-to-COS map \(\mathcal M_m\) is applied locally to the physical frequency
  leg of each core, producing the COS-coefficient core \(B_m\). (c) For
  arithmetic baskets, the coefficient train is contracted with the
  one-dimensional factors \(J_m(\cdot,u)\), yielding the basket characteristic
  function \(\varphi_H(u)\). The product in panel (c) is an ordered matrix-chain
  product over TT ranks, with boundary ranks equal to one.}
  \label{fig:tn-cos-tt-chf-workflow}
\end{figure}

Recall from \eqref{eq:basket-chf-from-cos} that
\[
  \varphi_H(u)
  \approx
  \sum_{k_1=0}^{N_1-1}{}'\cdots
  \sum_{k_d=0}^{N_d-1}{}'
  c_{\mathbf{k}}
  \prod_{m=1}^d J_m(k_m,u).
\]
Substituting the TT approximation \eqref{eq:cos-coeff-tt} gives
\[
  \varphi_H(u)
  \approx
  \sum_{k_1=0}^{N_1-1}{}'\cdots
  \sum_{k_d=0}^{N_d-1}{}'
  B_1(k_1)\cdots B_d(k_d)
  \prod_{m=1}^d J_m(k_m,u).
\]
For fixed \(u\), the product \(\prod_{m=1}^d J_m(k_m,u)\) is a separable
rank-one tensor over the COS multi-index \(\mathbf{k}\). Since each
\(J_m(k_m,u)\) is scalar, these factors can be absorbed into the corewise
sums, giving the sequential TT contraction
\begin{equation}
  \varphi_H(u)
  \approx
  \left(
    \sum_{k=0}^{N_1-1}{}'
    B_1(k)J_1(k,u)
  \right)
  \cdots
  \left(
    \sum_{k=0}^{N_d-1}{}'
    B_d(k)J_d(k,u)
  \right).
  \label{eq:basket-chf-tt-contraction}
\end{equation}
The tensor operations in
\eqref{eq:tt-chf}--\eqref{eq:basket-chf-tt-contraction} are summarized
diagrammatically in Figure~\ref{fig:tn-cos-tt-chf-workflow}. The TT-cross step introduces TT ranks between neighboring cores,
the CHF-to-COS maps act only on the physical frequency legs, and the arithmetic-basket evaluation contracts
the resulting COS-mode legs with the one-dimensional factors \(J_m(\cdot,u)\).

At this point the arithmetic-basket branch has all inputs needed for scalar COS
pricing: \(\varphi_H\) is obtained by TT contraction, and the remaining steps
are one-dimensional.

\section{Pricing from the COS Coefficient Tensor}
\label{sec:pricing-from-cos-coef}
This section explains how the coefficient tensor \(\mathcal{C}\), represented
in TT form by \eqref{eq:cos-coeff-tt}, is used after it has been constructed.
Arithmetic baskets are priced through the scalar basket density, while min/max
payoffs use rectangular probabilities of the joint log-price law.

For arithmetic baskets, the construction can be summarized as follows:
\[
\boxed{
  \varphi_X
  \longrightarrow
  \Phi\approx G_1\cdots G_d
  \longrightarrow
  \mathcal{C}\approx B_1\cdots B_d
  \longrightarrow
  \varphi_H
  \longrightarrow
  f_H
  \longrightarrow
  C(K),\,P(K)
}
\]
The diagram starts with the joint characteristic function
\(\varphi_X\).  The method samples \(\varphi_X\) on a tensor-product frequency
grid, compresses the sampled tensor \(\Phi\) in TT form, and converts the
compressed object into the COS coefficient tensor \(\mathcal{C}\).  For
arithmetic baskets, \(\mathcal{C}\) is then contracted to compute the scalar
basket characteristic function \(\varphi_H\).  The remaining steps are
one-dimensional: reconstruct \(f_H\) from \(\varphi_H\), then integrate \(f_H\)
against the call or put payoff.

Thus, after the joint COS coefficient tensor \(\mathcal{C}\) has been built,
arithmetic-basket pricing is one-dimensional: the TT contraction
\eqref{eq:basket-chf-tt-contraction} supplies \(\varphi_H(u_n)\), and the
scalar COS reconstruction \eqref{eq:basket-density-cos-reconstruction} and
prefix formulas \eqref{eq:basket-call-prefix}--\eqref{eq:basket-put-prefix}
give all requested strikes.  This is the strike-reuse mechanism measured in
Section~\ref{sec:multi-strike-reuse}.

The same coefficient tensor \(\mathcal{C}\) can also be used without
projecting to the basket density \(f_H\).  This is the route used for min/max
payoffs, where prices are obtained from rectangle probabilities of the joint
log-price law.

\subsection{Min/Max Payoffs from Joint COS--TT Coefficients}
\label{sec:minmax-cos-tt-derivation}
For min/max payoffs, we do not project the joint law to a scalar basket
density. Instead, we work directly with the COS coefficient tensor
\(\mathcal{C}=(c_{\mathbf{k}})\) of the joint log-price density \(f_X\). The
key observation is that tail events for the minimum and maximum become
coordinatewise conditions on the log-prices.

Let \(Y_T^{\min}=\min_{1\leq m\leq d}S_m(T)\).
The discounted call price is
\begin{equation}
  C_{\min}(K)
  =
  e^{-rT}\E\!\left[(Y_T^{\min}-K)^+\right].
  \label{eq:min-call-expectation}
\end{equation}
Since \(S_m(T)=e^{X_m}\), the same price can be written as
\[
  C_{\min}(K)
  =
  e^{-rT}
  \int_{\mathbb{R}^d}
  \left(\min_{1\leq m\leq d}e^{x_m}-K\right)^+
  f_X(x_1,\ldots,x_d)\,dx_1\cdots dx_d .
\]
The payoff \((\min_m e^{x_m}-K)^+\) is not itself a product of
one-dimensional factors, so the integral above does not directly reduce to a
separated tensor contraction.  A low-cost TT contraction arises once the
problem is rewritten in terms of tail probabilities, because the resulting
integration regions are Cartesian products. The tail representation below
supplies this separated structure.

For any
nonnegative random variable \(Y\),
\[
  (Y-K)^+
  =
  \int_K^\infty \mathbf{1}_{\{Y>y\}}\,dy .
\]
Taking expectations and interchanging expectation with the nonnegative integral
gives
\[
  \E[(Y-K)^+]
  =
  \int_K^\infty \mathbb{P}(Y>y)\,dy .
\]
This identity is useful because it replaces the nonseparable payoff
\((Y_T^{\min}-K)^+\) by a one-dimensional integral of tail probabilities.
Using the identity with \(Y=Y_T^{\min}\) in
\eqref{eq:min-call-expectation}, we obtain
\begin{equation}
  C_{\min}(K)
  =
  e^{-rT}
  \int_K^\infty Q_{\min}^{>}(y)\,dy,
  \qquad
  Q_{\min}^{>}(y)
  =
  \mathbb{P}\!\left(Y_T^{\min}>y\right).
  \label{eq:min-call-tail-integral}
\end{equation}
For each level \(y\), the inner problem is to compute the probability that all
assets finish above \(y\). In log-price variables,
\[
  Q_{\min}^{>}(y)
  =
  \mathbb{P}(X_m>\log y,\ m=1,\ldots,d).
\]
Equivalently, using the joint density \(f_X\) of \(X_T\),
\[
  Q_{\min}^{>}(y)
  =
  \int_{\log y}^{\infty}\cdots\int_{\log y}^{\infty}
  f_X(x_1,\ldots,x_d)\,dx_1\cdots dx_d .
\]
This is still a \(d\)-dimensional integral, but it has a much more useful shape
than the original payoff integral: each coordinate is integrated over a simple
upper-tail interval.

The COS approximation does not live on all of \(\mathbb{R}^d\), but on the
finite truncation box \(D\) from \eqref{eq:log-price-truncation-box}.  For
\(y>0\), intersect the coordinatewise upper-tail condition \(X_m>\log y\) with
the COS truncation interval \([a_m,b_m]\).  The resulting interval is
\([\log y,\infty)\cap [a_m,b_m]=[\ell_m(y),b_m]\), where \({\ell_m(y)=\min\{\max\{\log y,a_m\},b_m\}}\).  Consequently, the truncated
min-tail region is the Cartesian product
\(\prod_{m=1}^d [\ell_m(y),b_m]\).
The Cartesian-product form is what permits the corresponding probability to be
computed by contracting the joint COS coefficient tensor with one integrated
basis vector per dimension.

For fixed \(y\), the quantity approximated by the COS coefficient tensor is the
truncated min-tail probability
\[
  Q_{\min,D}^{>}(y)
  =
  \int_{\ell_1(y)}^{b_1}\cdots\int_{\ell_d(y)}^{b_d}
  f_X(x_1,\ldots,x_d)\,dx_1\cdots dx_d .
\]
We now evaluate the truncated min-tail probability using the same COS
coefficient representation introduced in \eqref{eq:joint-cos-expansion}.
Substituting the tensor-product COS expansion into \(Q_{\min,D}^{>}(y)\) gives
\[
  Q_{\min,D}^{>}(y)
  \approx
  \int_{\ell_1(y)}^{b_1}\cdots\int_{\ell_d(y)}^{b_d}
  \sum_{k_1=0}^{N_1-1}\cdots\sum_{k_d=0}^{N_d-1}{}'
  c_{k_1,\ldots,k_d}
  \prod_{m=1}^d \psi_{k_m}^{(m)}(x_m)
  \,dx_1\cdots dx_d .
\]
Exchanging the finite COS sums with the integrals, and using the fact that
\(\psi_{k_m}^{(m)}\) depends only on \(x_m\), yields
\[
  Q_{\min,D}^{>}(y)
  \approx
  \sum_{k_1=0}^{N_1-1}\cdots\sum_{k_d=0}^{N_d-1}{}'
  c_{k_1,\ldots,k_d}
  \prod_{m=1}^d
  \int_{\ell_m(y)}^{b_m}
  \psi_{k_m}^{(m)}(x_m)\,dx_m .
\]
This motivates the upper-tail integrated basis vector in dimension \(m\):
\[
  R_m(y;k)
  =
  \int_{\ell_m(y)}^{b_m}
  \psi_k^{(m)}(x)\,dx,
  \qquad k=0,\ldots,N_m-1 .
\]
Thus \(R_m(y;k)\) is the one-dimensional contribution of basis function \(k\)
to the \(m\)-th upper-tail interval.  With this notation,
\[
  Q_{\min,D}^{>}(y)
  \approx
  \sum_{k_1=0}^{N_1-1}\cdots\sum_{k_d=0}^{N_d-1}{}'
  c_{k_1,\ldots,k_d}
  \prod_{m=1}^d R_m(y;k_m).
\]

For later use with maximum payoffs, define the complementary lower-tail vector
\(A_m(y;k)=\int_{a_m}^{\ell_m(y)}\psi_k^{(m)}(x)\,dx\).  Writing
\(U_m(k)=\int_{a_m}^{b_m}\psi_k^{(m)}(x)\,dx\), we have
\(R_m(y)=U_m-A_m(y)\).  The clipping convention gives the correct endpoint
cases: \(R_m(y)=U_m\) if \(y\le e^{a_m}\), and \(R_m(y)=0\) if
\(y\ge e^{b_m}\).

Contracting the coefficient tensor with one-dimensional vectors means
\[
  \left\langle \mathcal{C}, v_1\otimes\cdots\otimes v_d\right\rangle
  :=
  \sum_{k_1=0}^{N_1-1}\cdots\sum_{k_d=0}^{N_d-1}
  c_{k_1,\ldots,k_d}
  \prod_{m=1}^d v_m(k_m).
\]
Here \(\mathcal{C}\) denotes the COS coefficient tensor, whereas
\(C_{\min}\) and \(C_{\max}\) denote call prices.
As in the tensor-product COS convention above, the prime weights are absorbed
into the vectors before contraction by halving each zero-mode entry. We keep
the same symbols for these weighted vectors.  With this convention,
\begin{equation}
  Q_{\min,D}^{>}(y)
  \approx
  \left\langle
  \mathcal{C},\,
  R_1(y)\otimes\cdots\otimes R_d(y)
  \right\rangle .
  \label{eq:min-call-rectangular-cos-contraction}
\end{equation}

For general \(d\), the direct contraction in
\eqref{eq:min-call-rectangular-cos-contraction} requires access to
\(\prod_{m=1}^d N_m\) coefficients.  With the TT representation
\eqref{eq:cos-coeff-tt}, the same quantity is computed as
\begin{equation}
  Q_{\min,D}^{>}(y)
  \approx
  \prod_{m=1}^d
  \left(
    \sum_{k_m=0}^{N_m-1}
    R_m(y;k_m)B_m(k_m)
  \right),
  \label{eq:min-call-rectangular-tt-contraction}
\end{equation}
where the boundary ranks \(r_0=r_d=1\) make the product scalar.  If
\(r=\max_m r_m\) and \(N=\max_m N_m\), the contraction costs
\(
  O(dNr^2)
\)
compared with \(O(N^d)\) for a
full tensor contraction.  The min-call price is therefore obtained as a
one-dimensional integral of low-rank rectangular-probability evaluations.

The other min/max payoffs use the same tail-integral idea as
\eqref{eq:min-call-tail-integral}, with different coordinatewise rectangles
and complementary tail identities.  For the maximum, the lower rectangle gives
\begin{equation}
  Q_{\max,D}^{\leq}(y)
  =
  \mathbb{P}_D\!\left(\max_{1\leq m\leq d}S_m(T)\leq y\right)
  \approx
  \left\langle
  \mathcal{C},\,
  A_1(y)\otimes\cdots\otimes A_d(y)
  \right\rangle .
  \label{eq:max-rectangular-cos-contraction}
\end{equation}
Here the subscript \(D\) emphasizes that the probability is computed inside
the COS truncation box.  Let
\(m_X=\left\langle\mathcal{C},U_1\otimes\cdots\otimes U_d\right\rangle\)
denote the recovered mass on \(D\).  The numerical pricing formulas use
complements with respect to \(m_X\):
\begin{align*}
  C_{\max}(K)
  &\approx
  e^{-rT}
  \int_K^{\max_m e^{b_m}}
  \left[m_X-Q_{\max,D}^{\leq}(y)\right]\,dy,\\
  P_{\min}(K)
  &\approx
  e^{-rT}
  \int_{\min_m e^{a_m}}^K
  \left[m_X-Q_{\min,D}^{>}(y)\right]\,dy,\\
  P_{\max}(K)
  &\approx
  e^{-rT}
  \int_{\max_m e^{a_m}}^K
  Q_{\max,D}^{\leq}(y)\,dy .
\end{align*}

After \(\mathcal{C}\) has been constructed, changing the strike changes only the
thresholds used in the integrated basis vectors and the outer one-dimensional
quadrature.  It does not require resampling the characteristic function or
recomputing the TT-cross approximation.  In computation, the rectangle
probabilities are clipped or checked against the interval \([0,m_X]\), and the
reported diagnostics include recovered mass, monotonicity across strikes, and
sensitivity to quadrature and truncation controls.

\subsection{Pricing Workflow and Diagnostics}
For a fixed model, payoff family, and numerical resolution, the mathematical
workflow is as follows.

\begin{enumerate}[label=\textbf{Step \arabic*.}, leftmargin=*]
  \item Choose the log-price truncation box \(D\), using
  \eqref{eq:log-price-truncation-box} in the reported experiments, together
  with frequency windows, quadrature nodes, COS orders, the TT-cross tolerance
  and rank cap, and the scalar basket interval \([a_H,b_H]\) when a basket
  density is reconstructed.

  \item Define the sampled-CHF entry oracle
  \[
    (j_1,\ldots,j_d)
    \mapsto
    \varphi_X(\omega_{1,j_1},\ldots,\omega_{d,j_d})
  \]
  and apply TT-cross to obtain \(\widehat{\Phi}\) in \eqref{eq:tt-chf}.

  \item Apply the one-dimensional maps \(\mathcal{M}_m\) in
  \eqref{eq:frequency-to-cos-map} to convert sampled-CHF TT cores into COS
  coefficient cores as in \eqref{eq:cos-coeff-core-transform}.

  \item For arithmetic baskets, compute the one-dimensional \(J_m(\cdot,u)\)
  vectors and absorb them into the coefficient-core contraction in
  \eqref{eq:basket-chf-tt-contraction},
  evaluate \(\varphi_H\) on the scalar COS frequency grid, reconstruct
  \(f_H\), and price all requested strikes from the prefix formulas
  \eqref{eq:basket-call-prefix}--\eqref{eq:basket-put-prefix}.

  \item For min/max payoffs, keep the joint COS coefficient TT, compute the
  required rectangle probabilities by contractions such as
  \eqref{eq:min-call-rectangular-tt-contraction} and
  \eqref{eq:max-rectangular-cos-contraction}, and price strikes from the
  resulting one-dimensional tail integrals.
\end{enumerate}

Diagnostics and summary statistics are computed from the same objects without
changing the pricing workflow.  The compression diagnostics compare held-out
entries of \(\widehat{\Phi}\) with direct characteristic-function evaluations.
The COS diagnostics include \(\varphi_H(0)\), density mass, first moment, and
put--call parity where applicable.  The numerical tables use these checks,
together with perturbation sweeps, independent reference-price deviations, TT
ranks, characteristic-function evaluation counts, and runtime components, to
assess the sensitivity of the reported prices to numerical controls.

\section{Selected Greek Computations}
\label{sec:selected-greeks}

The construction of the derivative characteristic-function tensor is
model-specific. Once its samples are available, their propagation through the
COS--TT--CHF maps and payoff contractions is independent of the model-specific
form of the characteristic function. Let \(\theta\) denote a market or model
input and let
\[
  \varphi_X(\omega;\theta)
  =
  \E_\theta\!\left[e^{\ii\omega^\top X_T}\right],
  \qquad \omega\in\mathbb{R}^d,
\]
be the joint characteristic function of the terminal log-price vector. Holding
the truncation box, quadrature nodes, COS orders, and payoff-contraction rules
fixed, a Greek is obtained by differentiating the sampled CHF tensor entrywise:
\[
  \partial_\theta \Phi[j_1,\ldots,j_d]
  =
  \partial_\theta
  \varphi_X(\omega_{1,j_1},\ldots,\omega_{d,j_d};\theta).
\]
The resulting tensor \(\partial_\theta\Phi\) is then passed through the same
linear COS--TT--CHF maps and payoff contractions as the price tensor.

The experiments below instantiate this construction for GBM arithmetic basket
options. We report the component spot-Delta vector with respect to
\(S_0=(S_1(0),\ldots,S_d(0))\), and the component Vega vector with respect to
\(\sigma=(\sigma_1,\ldots,\sigma_d)\). Both are computed by differentiating
the GBM characteristic-function representation before applying the
COS--TT--CHF maps.

Related tensor-train Fourier approaches compute Greeks by learning
parameter-dependent TT representations of Fourier pricing functions and then
either applying numerical differentiation operators to price-TT cores or
constructing separate analytical Greek TTs \citep{sakurai2025greeks}. In
contrast, we differentiate the characteristic-function representation at the
fixed model parameters and propagate the resulting derivative tensor through the
COS--TT--CHF maps.

Let
\(
  H=\sum_{i=1}^d w_iS_i(T)
\)
denote the terminal arithmetic basket, and let \(f_H\) be its density. For
nonnegative basket weights, the exact discounted call and put prices are
\[
  C(K)=e^{-rT}\int_K^\infty (h-K)f_H(h)\,dh,
  \qquad
  P(K)=e^{-rT}\int_0^K (K-h)f_H(h)\,dh.
\]
The numerical method applies the same payoff integrals to the reconstructed
one-dimensional COS approximation of \(f_H\) and to its propagated parameter
derivatives. The component Greeks below are therefore obtained by propagating
derivative tensors through the reconstruction and payoff-integration maps,
rather than by finite differences of reconstructed prices or densities. In the
definitions below, \(V\) denotes either \(C\) or \(P\) at the fixed strike
\(K\).
For spot sensitivity with respect to the initial-spot vector
\(S_0=(S_1(0),\ldots,S_d(0))\), define
\[
  \nabla_{S_0} V(K)
  =
  \left(
    \frac{\partial V(K)}{\partial S_1(0)},
    \ldots,
    \frac{\partial V(K)}{\partial S_d(0)}
  \right).
\]
For volatility sensitivity, the input volatility vector is
\(
  \sigma=(\sigma_1,\ldots,\sigma_d).
\)
Here \(\sigma_i\) denotes the volatility input in decimal units. The component
Vega vector is
\[
  \nabla_\sigma V(K)
  =
  \left(
    \frac{\partial V(K)}{\partial\sigma_1},
    \ldots,
    \frac{\partial V(K)}{\partial\sigma_d}
  \right).
\]
A scalar common-volatility Vega is obtained from the component vector by
scaling all volatilities by a common factor \(\eta\):
\[
  \left.
  \frac{\partial V(K;\eta\sigma)}{\partial \eta}
  \right|_{\eta=1}
  =
  \sum_{i=1}^d
  \sigma_i
  \frac{\partial V(K)}{\partial\sigma_i}.
\]

The simple single-core scaling used for GBM Delta below is model-specific. In
a general characteristic-function model, the derivative
characteristic-function tensor may instead be built as its own TT-cross object
and then propagated through the same COS--TT--CHF maps.

Both component constructions use derivatives of the GBM characteristic
function. Under correlated GBM,
\[
  \mu_i=\log S_i(0)+\left(r-\frac12\sigma_i^2\right)T,
  \qquad
  \Sigma_{ij}=\sigma_i\sigma_j\rho_{ij}T,
\]
and
\[
  \varphi_X(\omega)
  =
  \exp\left(
    \ii\omega^\top\mu-\frac12\omega^\top\Sigma\omega
  \right).
\]
For the spot derivative, only the mean vector depends on \(S_i(0)\). The
covariance matrix \(\Sigma\) is fixed. The chain rule gives
\[
  \frac{\partial \varphi_X(\omega)}{\partial S_i(0)}
  =
  \varphi_X(\omega)
  \frac{\partial}{\partial S_i(0)}
  \left(
    \ii\omega^\top\mu-\frac12\omega^\top\Sigma\omega
  \right).
\]
Moreover,
\[
  \frac{\partial\mu_j}{\partial S_i(0)}
  =
  \begin{cases}
    1/S_i(0), & j=i,\\
    0, & j\neq i.
  \end{cases}
\]
Therefore
\[
  \frac{\partial}{\partial S_i(0)}(\omega^\top\mu)
  =
  \sum_{j=1}^d
  \omega_j
  \frac{\partial\mu_j}{\partial S_i(0)}
  =
  \frac{\omega_i}{S_i(0)},
\]
and hence
\[
  \frac{\partial \varphi_X(\omega)}{\partial S_i(0)}
  =
  \frac{\ii\omega_i}{S_i(0)}\,\varphi_X(\omega).
\]
On the tensor frequency grid this factor depends only on coordinate \(i\).
Thus the component-Delta derivative tensor is obtained by scaling the
\(i\)-th TT core of the base characteristic-function tensor and then applying
the same COS--TT--CHF maps and payoff contractions as in pricing. In the
reported GBM tests in Table~\ref{tab:greeks}, this single-core scaling leaves
the observed TT ranks unchanged.

Differentiating the characteristic function with respect to \(\sigma_k\) gives
\[
  \frac{\partial \varphi_X(\omega)}{\partial \sigma_k}
  =
  \left[
    -\ii T\sigma_k\omega_k
    -T\omega_k\sum_{j=1}^d \rho_{kj}\sigma_j\omega_j
  \right]\varphi_X(\omega).
\]
The component Vega derivative tensor is propagated through the same linear maps
as the component Delta tensor. Before rounding, the algebraic construction
produces higher TT ranks. Table~\ref{tab:greeks} reports the maximum unrounded
rank. The derivative tensor is subsequently rounded with rank cap \(16\).
Common volatility-scale sensitivities are directional aggregates of the
component Vega vector.

\begin{figure}[t]
\centering
\includegraphics[width=0.92\textwidth]{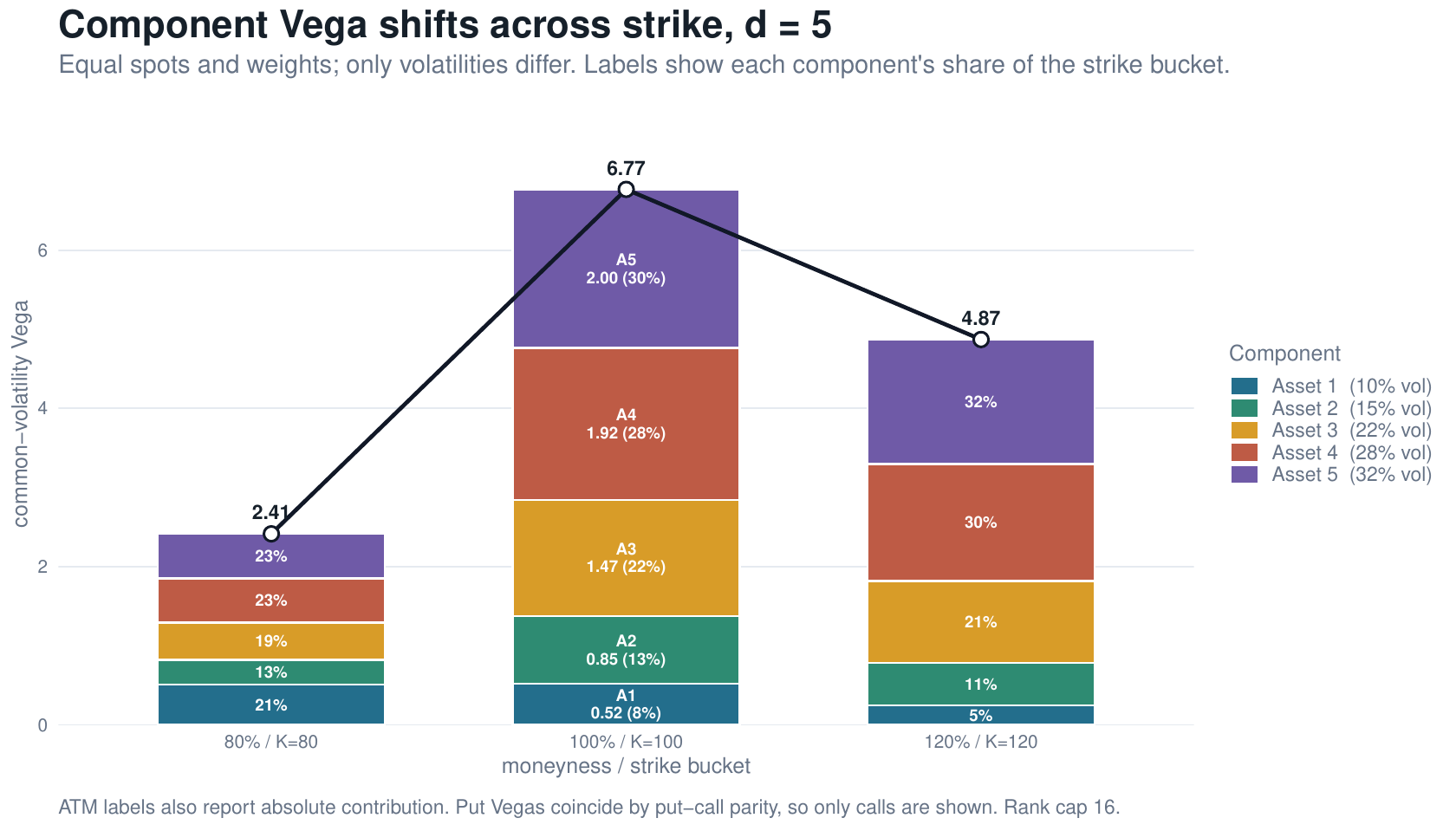}
\caption{Component Vega decomposition for a stylized five-asset GBM basket call.
The basket has \(S_i(0)=100\), equal weights \(w_i=0.2\), \(T=1\),
\(r=2\%\), \(K\in\{80,100,120\}\), volatilities
\(\sigma=(10,15,22,28,32)\%\), and Toeplitz correlation
\(\rho_{ij}=0.7^{|i-j|}\). Bars show the volatility-weighted component
contributions \(\sigma_i\,\partial C(K)/\partial\sigma_i\), which sum to the
common volatility-scale Vega
\(\partial C(K;\eta\sigma)/\partial\eta|_{\eta=1}\). Labels report each
asset's percentage contribution at the corresponding strike.}
\label{fig:component-vega}
\end{figure}

Figure~\ref{fig:component-vega} illustrates the common-volatility decomposition
for a stylized five-name basket. In this example the basket weights and initial
spots are identical, so the variation in the decomposition comes from the
volatility structure and the strike. At \(K=80\), the common-volatility Vega is
relatively evenly shared across names. Moving toward and above the money, the
contribution shifts toward the higher-volatility constituents: assets 4 and 5
together account for about \(58.8\%\) of the common-volatility Vega at
\(K=120\). The scalar aggregate gives the total common-volatility Vega, whereas
the component vector shows how that total is distributed across assets.

To check that the differentiated characteristic-function tensors produce the
intended first-order sensitivities, Table~\ref{tab:greeks} reports a focused
GBM diagnostic. This table is included here because it validates the Greek
construction itself. The broader pricing benchmarks are reported later.

The comparison with QMC is not intended to suggest that QMC cannot compute
Greeks. Simulation-based Greeks are available, but their accuracy depends
strongly on the estimator: finite-difference QMC Greeks require a bump-size
choice and can amplify simulation noise, while pathwise and conditional
estimators require payoff smoothness or additional smoothing. COS--TT--CHF
avoids these simulation-specific issues for the reported first-order Greeks:
the sensitivities are obtained by propagating differentiated
characteristic-function tensors through the same linear maps used for pricing,
so the relevant numerical controls are the transform, TT approximation, and
payoff-contraction diagnostics already used for prices.

For component Delta, the QMC reference uses the GBM pathwise identities
\[
  \Delta_i^C(K)
  =
  e^{-rT}
  \E\left[
    w_i\frac{S_i(T)}{S_i(0)}
    \mathbf{1}_{H>K}
  \right],
  \qquad
  \Delta_i^P(K)
  =
  -e^{-rT}
  \E\left[
    w_i\frac{S_i(T)}{S_i(0)}
    \mathbf{1}_{H<K}
  \right].
\]
For component Vega, QMC uses central common-random-number finite differences
asset by asset. Let \(\sigma^{(i,+)}\) and \(\sigma^{(i,-)}\) denote the
volatility vectors obtained from \(\sigma\) by replacing \(\sigma_i\) with
\((1+h)\sigma_i\) and \((1-h)\sigma_i\), respectively. Then
\[
  \mathrm{Vega}^{\mathrm{QMC}}_i(K)
  \approx
  \frac{
  V_{\mathrm{QMC}}(K;\sigma^{(i,+)})
  -
  V_{\mathrm{QMC}}(K;\sigma^{(i,-)})
  }{2h\sigma_i},
  \qquad h=10^{-3},
\]
using the same randomized Sobol normal variates in the up and down evaluations.

\begin{table}[t]
\centering
\caption{Component Greek diagnostics for GBM arithmetic basket options.}
\label{tab:greeks}
\scriptsize
\setlength{\tabcolsep}{3pt}
\begin{tabular}{@{}rrrrrr@{}}
\toprule
\multicolumn{6}{l}{\textbf{Panel A: Analytic component spot Delta}}\\
\midrule
\(d\) & Base (s) & Delta (s) & Rank & Max err./\(w_i\) & QMC half/\(w_i\) \\
\midrule
2 & 0.014 & 0.012 & 13 & \(1.041\times10^{-4}\) & \(6.433\times10^{-5}\) \\
5 & 0.135 & 0.074 & 17 & \(8.769\times10^{-5}\) & \(3.418\times10^{-4}\) \\
10 & 2.472 & 0.249 & 18 & \(3.345\times10^{-4}\) & \(4.978\times10^{-4}\) \\
20 & 8.369 & 0.855 & 18 & \(2.484\times10^{-4}\) & \(5.549\times10^{-4}\) \\
\bottomrule
\end{tabular}

\medskip
\begin{tabular}{@{}rrrrrr@{}}
\toprule
\multicolumn{6}{l}{\textbf{Panel B: Analytic component Vega with rank control}}\\
\midrule
\(d\) & Base (s) & Vega (s) & Raw rank & Max err./\(w_i\) & QMC half/\(w_i\) \\
\midrule
2 & 0.084 & 0.013 & 39 & 0.0275 & 0.0065 \\
5 & 0.230 & 0.595 & 102 & 0.0395 & 0.0381 \\
10 & 2.860 & 5.353 & 198 & 0.0592 & 0.0522 \\
20 & 9.399 & 46.434 & 378 & 0.0805 & 0.0797 \\
\bottomrule
\end{tabular}
\par\smallskip
{\footnotesize Panel A reports component spot Delta and Panel B reports
component Vega. Panel B uses TT rounding with derivative rank cap 16. In Panel
A, the component-Delta rank equals the base rank. Error columns compare
COS--TT--CHF component Greeks with QMC references. Since the basket is equally
weighted, \(w_i=1/d\), errors and QMC half-widths are divided by \(w_i\). The
reported maximum error is the worst normalized component error across all
assets, strikes, and both call and put options. Each QMC reference uses 32
independently scrambled Sobol point sets with \(2^{15}\) points per scramble,
for \(2^{20}\) points per dimension. The reported 95\%
confidence half-widths are computed from variation across scrambles.}
\end{table}

Panel A shows that component Delta keeps the base TT rank in the tested
dimensions. With the reported component-Delta settings, the largest normalized
component error is \(3.345\times10^{-4}\), and the \(d=20\)
component-Delta post-processing time is \(0.855\) seconds after the base
construction. Panel B shows that component Vega is more expensive because the
derivative construction increases TT ranks. With rank cap \(16\), the maximum
normalized component-Vega error is \(8.05\times10^{-2}\). The diagnostic
supports the use of the differentiated-CHF route for first-order component
Greeks. Broader Greek-surface studies are a natural extension beyond the
pricing benchmarks reported below.

Other Greeks can be derived by differentiating the characteristic function
further, when the required derivatives exist and can be represented accurately.
For example, spot Gamma consists of the second derivatives
\(
  \frac{\partial^2 V}{\partial S_i(0)\,\partial S_j(0)}.
\)

\section{Numerical Error, Diagnostics, and Cost}
The preceding sections define the pricing and sensitivity maps. We now separate
the numerical errors that enter those maps from the validation errors measured
against reference prices.

\subsection{Error Sources}
COS--TT--CHF introduces several approximation layers, and these layers are
controlled by different numerical parameters. For notational economy, write
\(W\), \(M\), and \(N\) for the common frequency half-width, number of
frequency nodes, and COS order used in the experiments. The construction also
allows coordinate-dependent values.  The node distribution is also treated as
a numerical control.  This flexibility is useful because different
characteristic functions have different decay and oscillation patterns: GBM,
VG, NIG, and affine stochastic-volatility examples need not be resolved
equally well by the same frequency window and node distribution.  The pricing
workflow can be summarized by the object chain
\[
  \varphi_X
  \longrightarrow
  \Phi^{W,M}
  \longrightarrow
  \widehat{\Phi}^{W,M}
  \longrightarrow
  \widehat c^{N}
  \longrightarrow
  \widehat V .
\]
Here \(\varphi_X\) is the model characteristic function of the terminal
log-price vector. The tensor \(\Phi^{W,M}\) contains its values on the
tensor-product frequency grid, \(\widehat{\Phi}^{W,M}\) is the TT-cross
approximation to this sampled tensor, and \(\widehat c^N\) is the compressed
COS coefficient object after the CHF-to-COS transforms with \(N\) retained
modes. Thus the numerical error can enter through the truncation domain and
frequency window, the finite frequency grid, the TT-cross approximation, the
finite COS order, and the final contractions and one-dimensional integrations.

The reported value \(\widehat V\) is obtained after payoff-specific
post-processing. Arithmetic baskets use basket-density reconstruction followed
by one-dimensional payoff integration. Min/max payoffs use
rectangle-probability contractions followed by tail-integral pricing formulas.
We use the decomposition above as an accounting device for numerical controls
and diagnostics, not as a sharp a priori error bound.

Reference values are used only for validation. An observed discrepancy
\(|\widehat V - V_{\mathrm{ref}}|\) measures agreement with the chosen
reference, not the unknown error \(|\widehat V - V|\) itself. QMC references
therefore include confidence intervals and, when relevant, discretization-
refinement diagnostics. High-resolution COS--TT--CHF references are treated as
refinement checks.

The same separation applies to the selected Greeks. Component Delta and
component Vega are obtained by differentiating the GBM characteristic function
with respect to each initial-spot or volatility component and propagating the
derivative objects through the same linear COS--TT--CHF maps. Their validation
errors are measured against pathwise or finite-difference QMC references for
the individual components.

\subsection{Diagnostic Checks}
The reported tables include diagnostics computed from the same objects used for
pricing. These diagnostics are not independent error bounds. They check
identities and shape constraints that should hold for the reconstructed density
and the resulting price grid.

For arithmetic baskets, the reconstructed density \(f_H\) should have unit mass
and, when the basket mean is known, the correct first moment. We therefore
monitor
\[
  \epsilon_{\mathrm{mass}}
  =
  \left|\int f_H(h)\,dh - 1\right|,
  \qquad
  \epsilon_{\mathrm{mean}}
  =
  \left|\int h f_H(h)\,dh - \E[H]\right|.
\]
These quantities test the shared density used to price all strikes in the
basket price grid.

The held-out CHF-entry checks are especially important for COS--TT--CHF,
because an accurate payoff calculation requires the compressed frequency-domain
object to resolve the oscillatory structure of \(\varphi_X\) before the
CHF-to-COS transform is applied.

We also monitor pricing identities and shape constraints. For basket calls and
puts, put--call parity gives
\[
  C(K)-P(K)=e^{-rT}(\E[H]-K),
\]
and the tables report the corresponding parity residual. Across a strike grid,
call prices should be nonincreasing, put prices should be nondecreasing, and
prices should remain nonnegative. For min/max payoffs, the analogous checks are
applied to rectangle probabilities and tail probabilities before the final
tail-integral pricing step.

These diagnostics are interpreted together with reference-price deviations and
one-at-a-time perturbation sweeps. Small residuals do not prove that a price is
exact, but large residuals suggest which part of the numerical pipeline may
need refinement.

\subsection{Sensitivity to Numerical Controls}
The following one-at-a-time sweep is a local sensitivity diagnostic, not the
acceptance test for the final VG price surface. It retains the direct call and
put reconstructions so that mass and parity sensitivity remain visible. One
control is varied around a baseline configuration while the remaining controls
are held fixed.

Table~\ref{tab:vg-numerical-stability} reports the sweep for the canonical
\(d=20\) VG basket case from Table~\ref{tab:benchmark-market-specs}. The sweep
varies the TT rank cap, the TT-cross seed, the frequency half-width \(W\), the
number of frequency nodes \(M\), and the COS order \(N\). The baseline row
uses rank cap \(20\), \(W=24\), \(M=48\), \(N=64\), and seed \(42\). The
columns ``Ref. dev.'' and ``Baseline change'' report the maximum call/put
deviation from a higher-resolution internal reference and from the baseline
row, respectively. The mass and parity columns are the diagnostics from the
previous subsection.

\begin{table}[htbp]
\centering
\caption{One-at-a-time sensitivity sweep for the VG \(d=20\) COS--TT--CHF basket case.}
\label{tab:vg-numerical-stability}
\scriptsize
\begin{tabular}{llrrrrrr}
\toprule
Control & Setting & Call & Put & Ref. dev. & Baseline change & Mass err. & Parity resid. \\
\midrule
baseline & default & 6.816795 & 3.865618 & \(4.479\times10^{-3}\) & \(0.000\times10^{0}\) & \(1.272\times10^{-3}\) & \(4.269\times10^{-3}\) \\
rank cap & 16 & 6.813691 & 3.863179 & \(7.330\times10^{-3}\) & \(3.104\times10^{-3}\) & \(1.517\times10^{-3}\) & \(4.935\times10^{-3}\) \\
rank cap & 24 & 6.817420 & 3.867929 & \(3.601\times10^{-3}\) & \(2.311\times10^{-3}\) & \(1.236\times10^{-3}\) & \(5.955\times10^{-3}\) \\
frequency width & \(W=20\) & 6.818375 & 3.871242 & \(2.646\times10^{-3}\) & \(5.625\times10^{-3}\) & \(1.108\times10^{-3}\) & \(8.314\times10^{-3}\) \\
frequency width & \(W=28\) & 6.800936 & 3.860651 & \(2.009\times10^{-2}\) & \(1.586\times10^{-2}\) & \(2.738\times10^{-3}\) & \(1.516\times10^{-2}\) \\
nodes/axis & \(M=40\) & 6.836835 & 3.867768 & \(1.581\times10^{-2}\) & \(2.004\times10^{-2}\) & \(1.301\times10^{-3}\) & \(1.362\times10^{-2}\) \\
nodes/axis & \(M=56\) & 6.821649 & 3.869674 & \(6.281\times10^{-4}\) & \(4.854\times10^{-3}\) & \(8.574\times10^{-4}\) & \(3.471\times10^{-3}\) \\
COS order & \(N=48\) & 6.817111 & 3.865646 & \(4.450\times10^{-3}\) & \(3.153\times10^{-4}\) & \(1.270\times10^{-3}\) & \(3.982\times10^{-3}\) \\
COS order & \(N=96\) & 6.816795 & 3.865618 & \(4.479\times10^{-3}\) & \(1.716\times10^{-8}\) & \(1.272\times10^{-3}\) & \(4.269\times10^{-3}\) \\
TT-cross seed & 1729 & 6.817725 & 3.863903 & \(6.193\times10^{-3}\) & \(1.714\times10^{-3}\) & \(1.403\times10^{-3}\) & \(1.625\times10^{-3}\) \\
TT-cross seed & 1910 & 6.817667 & 3.867191 & \(3.355\times10^{-3}\) & \(1.573\times10^{-3}\) & \(1.242\times10^{-3}\) & \(4.971\times10^{-3}\) \\
\bottomrule
\end{tabular}
\end{table}

The sweep indicates that the local variation is not primarily driven by
TT-cross seed randomness. Seed changes, moderate rank-cap changes, and the
final COS-order perturbation produce smaller shifts than the stressed
frequency-window and node-count rows. Because \(W\) and \(M\) are perturbed
separately, the frequency rows are interpreted as sensitivity diagnostics
rather than as a joint frequency-grid convergence sequence. Accuracy of the
practical VG price surface is assessed separately using parity-consistent
prices and independent randomized Sobol QMC in
Section~\ref{sec:multi-strike-reuse}. The analogous GBM sensitivity sweep is
reported in Appendix~\ref{app:additional-stability-sweeps}.

\subsection{Cost Model}
The cost reduction in COS--TT--CHF is conditional on the sampled CHF and COS
coefficient tensors having moderate TT ranks. Without compression, the sampled
characteristic-function tensor has \(\prod_{m=1}^d M_m\) entries, and the COS
coefficient tensor has \(\prod_{m=1}^d N_m\) entries. A TT representation with
ranks \(r_0=r_d=1\) stores
\[
  \sum_{m=1}^d n_m r_{m-1}r_m
\]
numbers for a tensor with mode sizes \(n_m\). If \(n_m=n\) and the intermediate
TT ranks are bounded by \(r\), this storage
is \(O(dnr^2)\), rather than \(O(n^d)\) for a full tensor.

TT-cross constructs the sampled-CHF tensor from selected entries, so the cost
of the compression step is measured by the number of characteristic-function
evaluations and by the resulting ranks. The runtime and diagnostic tables
therefore report ranks and characteristic-function evaluation counts together
with wall-clock times. Dimension alone is not an adequate proxy for cost:
a higher-dimensional Gaussian example can be cheaper than a lower-dimensional
non-Gaussian example if the effective TT ranks are smaller.

After the sampled CHF tensor has been compressed, the CHF-to-COS step transforms
each TT core separately. The remaining pricing cost depends on the payoff
family. Basket pricing reconstructs a one-dimensional basket density and reuses
cumulative integrals across strikes. Min/max pricing uses the joint coefficient
TT to compute rectangle probabilities and then prices strikes through tail
integrals. The expensive step is therefore the one-time construction of the
compressed representation. Strike grids and selected post-processing quantities
reuse that representation.

These cost statements are conditional. In the worst case, TT ranks can grow
rapidly with dimension. For this reason, the numerical tables report ranks,
characteristic-function evaluation counts, runtimes, and diagnostic residuals
rather than dimension alone.

\section{Benchmark Protocol}
With the numerical objects and diagnostics fixed, we now specify the benchmark
protocols used in the reported experiments.

\subsection{Comparator Methods and Scope of Comparison}
The comparison methods used below reduce the dimensional cost of Fourier
pricing at different points in the numerical pipeline. We summarize them before
the benchmark tables because the reported timings and reference values are not
all measurements of the same numerical object.

\citet{bayer2023} formulate multi-asset option prices as damped Fourier
integrals involving the model characteristic function and the Fourier
transform of the payoff. Their main numerical controls
are the damping vector, which is chosen to improve integrability and smoothness
of the Fourier integrand, and the quadrature rule used to evaluate the
resulting \(d\)-dimensional integral. The method combines optimized damping
with dimension-adaptive sparse-grid quadrature. This makes it a useful
published reference for medium-dimensional basket prices under GBM, VG, and
NIG dynamics. The later Fourier-RQMC variant replaces deterministic sparse-grid
quadrature by randomized QMC after a domain transformation from
\(\mathbb{R}^d\) to \([0,1]^d\) \citep{bayer2024rqmc}. In this paper, the
Bayer et al. rows in Table~\ref{tab:medium-dimensional-validation} use their
published OD+ASGQ convention. The reported OD+ASGQ runtimes are literature
context rather than same-machine timings.

\citet{junike2025} develop classical and damped multidimensional COS methods for
recovering distribution functions, moments, and option prices from a
characteristic function. The method truncates the density to
a finite hypercube and approximates it by a multidimensional Fourier-cosine
series. For payoffs such as arithmetic basket puts, their damped COS-iv variant
uses exponential damping and the Fourier transform of the payoff to avoid
numerically computing payoff cosine coefficients by a discrete cosine
transform. This is the closest direct-COS comparator to COS--TT--CHF. The
important distinction is that COS-iv remains a direct
tensor-product COS calculation, while COS--TT--CHF compresses the sampled
characteristic-function tensor before forming the COS coefficient object. The
local tuned COS-iv baseline therefore provides same-machine evidence for the
crossover between direct tensor COS and the compressed representation. In
Table~\ref{tab:medium-dimensional-validation}, the Junike--Stier source rows
use their published basket-put cases for price validation and a local tuned
COS-iv implementation as the direct tensor-COS runtime baseline.

\citet{kastoryano2022} use tensor networks in a different Fourier
representation. They discretize a Lewis-type Fourier
pricing formula as an inner product between two high-dimensional tensors: one
containing values of the characteristic function and one containing the Fourier
transform of the payoff. Both tensors are approximated in matrix-product-state
or tensor-train form using TT-cross, and the option price is obtained by
contracting the two compressed objects. This is the closest tensor-network
comparator, especially for the call-on-minimum payoff. The distinction is where
the reusable object lives: Kastoryano--Pancotti compress a payoff-specific
Fourier integrand, whereas COS--TT--CHF builds a compressed COS coefficient
representation from the characteristic function and then applies payoff maps to
that representation. Accordingly, the KP TF timing column in
Table~\ref{tab:kp-min-option-fast-target} is literature context rather than a
same-machine runtime comparison.

\begin{table}[htbp]
\centering
\caption{Comparator methods and the intended scope of comparison.}
\label{tab:comparator-methods}
\scriptsize
\setlength{\tabcolsep}{3pt}
\begin{tabular}{@{}>{\raggedright\arraybackslash}p{0.14\textwidth}
                  >{\raggedright\arraybackslash}p{0.23\textwidth}
                  >{\raggedright\arraybackslash}p{0.22\textwidth}
                  >{\raggedright\arraybackslash}p{0.18\textwidth}
                  >{\raggedright\arraybackslash}p{0.15\textwidth}@{}}
\toprule
Method & Numerical object & Dimension strategy & Role in this paper & Scope note \\
\midrule
OD+ASGQ \citep{bayer2023} &
Damped Fourier integral over the pricing transform &
Optimized damping plus dimension-adaptive sparse-grid quadrature &
Published basket-price validation and runtime context &
Source-paper timings are not same-machine. \\
COS-iv \citep{junike2025} &
Damped COS sum over density and payoff coefficients &
Direct tensor-product COS modes with damping, truncation-range, and COS-order
choices &
Same-machine direct-COS baseline for basket puts &
Tests compression against full tensor COS. \\
Tensor-Fourier \citep{kastoryano2022} &
Compressed Fourier inner product between characteristic-function and payoff
tensors &
MPS/TT compression of both Fourier tensors by TT-cross &
Tensor-network min-option bridge &
Payoff-specific Fourier-integrand compression. \\
COS--TT--CHF &
Compressed COS coefficient tensor built from CHF samples &
TT-cross on CHF samples followed by corewise CHF-to-COS transforms and
payoff-specific contractions &
Method studied in this paper &
Reuse is valuable when effective TT ranks stay moderate. \\
\bottomrule
\end{tabular}
\end{table}

\subsection{Benchmark Families and Market Specifications}
The numerical study uses two types of benchmarks: generated benchmark families
used to test scaling and reuse, and published benchmark conventions used for
validation or comparison. For the generated basket experiments, unless
otherwise stated, the basket has \(S_i(0)=100\), equal weights \(w_i=1/d\),
maturity \(T=1\), and arithmetic basket call/put payoffs. In this section,
\(r\) denotes the risk-free rate.
Table~\ref{tab:benchmark-market-specs} fixes only market, payoff, dimension,
and strike conventions. Numerical controls, reference methods, tolerances, and
diagnostic quantities are reported with the corresponding result or appendix
table. The Bayer et al. and Junike--Stier validation cases retain their source-paper
conventions and are summarized separately in
Table~\ref{tab:medium-dimensional-benchmark-conventions}.
For the heterogeneous asset-loading rules below,
\(x_i=(i-1)/(d-1)\).

\newcommand{\benchmarkdk}[2]{%
  \begin{tabular}[t]{@{}l@{\;=\;}>{\raggedright\arraybackslash}p{0.155\textwidth}@{}}
    \(d\) & #1\\
    \(K\) & #2
  \end{tabular}%
}

\begin{table}[htbp]
\centering
\caption{Market specifications used by the main numerical experiments.}
\label{tab:benchmark-market-specs}
\scriptsize
\setlength{\tabcolsep}{3pt}
\begin{tabular}{@{}>{\raggedright\arraybackslash}p{0.18\textwidth}
                  >{\raggedright\arraybackslash}p{0.22\textwidth}
                  >{\raggedright\arraybackslash}p{0.52\textwidth}@{}}
\toprule
Experiment & Dimensions and strikes & Market, model, and payoff convention \\
\midrule
VG sensitivity and strike reuse &
\benchmarkdk{\(20\)}{\(100\) for sensitivity and \(80{:}0.5{:}120\) for reuse} &
\(r=0.03\), \(\theta=-0.30\), \(\nu=0.10\),
\(\sigma_i=0.20+0.20x_i+0.05\sin(\pi x_i)\),
\(\rho_{ij}=0.35^{|i-j|}\). \\
High-dimensional GBM and Greeks &
\benchmarkdk{pricing: \(10,15,20,30\), and Greeks: \(2,5,10,20\)}
{pricing: \(80,90,100,110,120\), and Greeks: \(80,100,120\)} &
\(r=0.02\),
\(\sigma_i=0.18+(0.30-0.18)x_i+0.015\sin(\pi x_i)\),
\(\rho_{ij}=0.7^{|i-j|}\). \\
Runtime-scaling VG/NIG &
\benchmarkdk{\(5,10,15,20\)}{\(80,90,100,110,120\)} &
VG uses the same heterogeneous VG specification as the sensitivity/reuse row.
NIG uses \(\alpha=20\), \(\beta_i=-2\), \(\delta=0.2\), \(r=0.03\), and
\(T=1\), with shape matrix \(\Delta_{ij}=0.35^{|i-j|}\). Initial spots are
\(S_i(0)=100\), and basket weights are \(w_i=1/d\). \\
Common-Heston stress test &
\benchmarkdk{\(2,5,10,20\)}{\(80,90,100,110,120\)} &
\(r=0.02\), \(\kappa=2\), \(\theta=v_0=0.04\), \(\xi=0.07\),
common leverage \(\beta_i=-0.25\), effective volatilities from 18\% to 30\%,
and residual correlation decay \(0.45\). \\
Kastoryano--Pancotti min-option bridge &
\benchmarkdk{\(2,3,4,5,10,15\), with extensions at \(20,30\)}{\(100\)} &
Correlated GBM call on the minimum with \(S_i(0)=K=100\), \(T=1\),
\(r=0.3\), common volatility \(\sigma=0.5\), and correlation matrix
\(\Sigma(\beta)=(\beta |+\rangle\langle+|+I)/(1+\beta)\), using
\(\beta=0.5\) for the primary bridge. \\
Literature validation &
Bayer et al. and Junike--Stier cases &
Published benchmark conventions, kept separate from the generated GBM/VG/NIG
specifications. \\
\bottomrule
\end{tabular}
\end{table}

The next sections apply these conventions to validation, scaling, reuse, and
Greek diagnostics.

\section{Validation Against Existing Benchmarks}
\subsection{Published Basket Benchmarks}
We next compare COS--TT--CHF against medium-dimensional arithmetic basket-put
benchmarks from the OD+ASGQ study of Bayer et al. and from Junike--Stier. These
cases test the full pricing workflow on published reference examples before
moving to the high-dimensional experiments.
Table~\ref{tab:medium-dimensional-validation} reports both
published-reference price reproduction and runtime context for these benchmark
cases. The source-specific market conventions are collected in
Appendix~\ref{app:benchmark-conventions}.

\begin{table}[htbp]
\centering
\caption{Medium-dimensional published-reference validation and runtime context.}
\label{tab:medium-dimensional-validation}
\scriptsize
\begin{tabular}{lllrrrcr}
\toprule
\multicolumn{8}{l}{\textbf{Panel A: Published-reference price validation}}\\
\midrule
Source & Case & Model & \(d\) & \(K\) & Ref. & COS--TT--CHF & Error \\
\midrule
Bayer et al. & B5 & GBM & 4 & 100 & 8.193000 & 8.192763 & \(2.895{\times}10^{-5}\) rel. \\
Bayer et al. & B6 & GBM & 4 & 100 & 11.301400 & 11.302065 & \(5.889{\times}10^{-5}\) rel. \\
Bayer et al. & B17 & VG & 4 & 100 & 8.944100 & 8.943911 & \(2.113{\times}10^{-5}\) rel. \\
Bayer et al. & B18 & VG & 4 & 100 & 11.227700 & 11.230308 & \(2.323{\times}10^{-4}\) rel. \\
Bayer et al. & B9 & GBM & 6 & 60 & 0.004100 & 0.004100 & \(2.263{\times}10^{-5}\) rel. \\
Bayer et al. & B10 & GBM & 6 & 60 & 0.012702 & 0.012706 & \(2.823{\times}10^{-4}\) rel. \\
Bayer et al. & B21 & VG & 6 & 60 & 0.169100 & 0.168933 & \(9.872{\times}10^{-4}\) rel. \\
Bayer et al. & B22 & VG & 6 & 60 & 0.046340 & 0.046539 & \(4.295{\times}10^{-3}\) rel. \\
Junike--Stier & JS-BS-d2 & BS/GBM & 2 & 100 & 6.906600 & 6.906905 & \(3.051{\times}10^{-4}\) abs. \\
Junike--Stier & JS-BS-d4 & BS/GBM & 4 & 100 & 6.305600 & 6.305960 & \(3.604{\times}10^{-4}\) abs. \\
Junike--Stier & JS-VG-d2 & VG & 2 & 100 & 5.595100 & 5.595137 & \(3.677{\times}10^{-5}\) abs. \\
Junike--Stier & JS-VG-d4 & VG & 4 & 100 & 3.969600 & 3.969705 & \(1.048{\times}10^{-4}\) abs. \\
\bottomrule
\end{tabular}

\medskip
\begin{tabular}{lllrlrcrr}
\toprule
\multicolumn{9}{l}{\textbf{Panel B: Runtime context}}\\
\midrule
Source & Case & Model & \(d\) & Baseline & Base (s) & COS--TT--CHF (s) & Max rank & CF evals\\
\midrule
Bayer et al. & B5 & GBM & 4 & OD+ASGQ (reported) & 7.80 & 0.076 & 1 & 700\\
Bayer et al. & B6 & GBM & 4 & OD+ASGQ (reported) & 2.73 & 0.097 & 1 & 840\\
Bayer et al. & B17 & VG & 4 & OD+ASGQ (reported) & 5.00 & 0.232 & 13 & 93900\\
Bayer et al. & B18 & VG & 4 & OD+ASGQ (reported) & 2.00 & 0.150 & 13 & 93900\\
Bayer et al. & B9 & GBM & 6 & OD+ASGQ (reported) & 2.00 & 0.088 & 1 & 1100\\
Bayer et al. & B10 & GBM & 6 & OD+ASGQ (reported) & 2.10 & 0.110 & 1 & 1760\\
Bayer et al. & B21 & VG & 6 & OD+ASGQ (reported) & 2.30 & 0.317 & 19 & 476000\\
Bayer et al. & B22 & VG & 6 & OD+ASGQ (reported) & 3.50 & 0.489 & 19 & 856800\\
Junike--Stier & JS-BS-d2 & BS/GBM & 2 & COS-iv (local tuned) & 0.001 & 0.029 & 17 & 9720\\
Junike--Stier & JS-BS-d4 & BS/GBM & 4 & COS-iv (local tuned) & 0.367 & 0.190 & 21 & 270208\\
Junike--Stier & JS-VG-d2 & VG & 2 & COS-iv (local tuned) & \(<10^{-3}\) & 0.022 & 9 & 2880\\
Junike--Stier & JS-VG-d4 & VG & 4 & COS-iv (local tuned) & 2.342 & 0.051 & 9 & 30272\\
\bottomrule
\end{tabular}
\begin{flushleft}
\footnotesize
In Panel A, Bayer et al. errors are relative errors and Junike--Stier errors are
absolute errors, following the corresponding source conventions.
For every Bayer et al. row, the COS--TT--CHF relative error is below the
corresponding OD+ASGQ relative error reported in their Table~4.6.
Junike--Stier rows use an absolute-error target of about \(10^{-2}\), which all
four rows meet. COS--TT--CHF times are local median
runtimes. Reported OD+ASGQ times are not hardware-normalized against the
local runs. For the Junike--Stier rows, the COS-iv baseline is a local tuned
run of the public implementation. The \(d=2\) local direct-COS timings are near
the measurement floor.
\end{flushleft}
\end{table}

Table~\ref{tab:medium-dimensional-validation} separates two forms of evidence.
The Bayer et al. rows are the main published-reference validation set: they cover
GBM and VG arithmetic basket puts in dimensions \(d=4\) and \(d=6\), and
the COS--TT--CHF relative error is below the corresponding reported OD+ASGQ
relative error in every row.  The
reported COS--TT--CHF runtimes are also below the published OD+ASGQ runtimes
in all Bayer et al. rows.  This comparison is encouraging but not
hardware-normalized: Bayer et al. report OD+ASGQ timings from a 72-core,
256 GB cluster node, whereas our COS--TT--CHF timings are local medians from
an Apple M3 Pro MacBook Pro with 11 CPU cores and 18 GB of memory.  Since
our machine is not stronger than the reported OD+ASGQ hardware, the lower
COS--TT--CHF times are still informative, but they should be read as
literature-context evidence rather than same-machine speedups.  The runtime
diagnostics are consistent with the tensor structure of the models.
Independent GBM rows compress to rank \(1\), while the VG rows require ranks
\(13\) and \(19\), increasing the number of characteristic-function samples.
The Bayer et al. bridge therefore shows that COS--TT--CHF reproduces established
medium-dimensional prices and that its cost tracks numerical rank rather than
nominal dimension alone.

The Junike--Stier rows provide the same-machine direct tensor-COS comparison.
Their published \(d=2,4\) arithmetic basket-put references are reproduced
within the stated absolute-error target, and the public COS-iv implementation
is run locally as a tuned baseline.  For each row, direct COS-iv is tuned to
the first tested configuration reaching the \(10^{-2}\) target.  This
accuracy-matched baseline is deliberately conservative: it is faster than the
published-parameter COS-iv run whenever a smaller COS grid already reaches
the target.  Against this tuned baseline, COS--TT--CHF is slower in the
timing-floor \(d=2\) rows, but faster in both \(d=4\) rows: \(0.190\) seconds
versus \(0.367\) seconds for BS/GBM and \(0.051\) seconds versus \(2.342\)
seconds for VG.

As additional same-machine context, we also ran a local COS-iv-style \(d=5\)
stress extension based on the Junike--Stier cases. This extension is not part
of the published benchmark table. It uses the same public COS-iv code and
scrambled Sobol references based on 16 independent scrambles with
\(2^{19}\) points per scramble, for \(2^{23}\) points
per model, to probe the next dimension. At \(N=20\), direct COS-iv took
\(55.7\) seconds for the BS/GBM case and \(115.3\) seconds for the VG case, while missing
the \(10^{-2}\) target.  COS--TT--CHF reached approximately
\(7{\times}10^{-5}\) absolute error on both cases in subsecond time.  Thus,
on the local COS-iv-style extension, COS--TT--CHF is both faster and more
accurate than the tested direct COS-iv configurations, illustrating where
tensor compression becomes decisive as the tensor-product grid grows.

\subsection{Kastoryano--Pancotti Min-Option Bridge}
\label{sec:kp-min-option-bridge}

The target payoff in this bridge is the call on the minimum,
\(C_{\min}(K)\), matching equation~(11) of \citet{kastoryano2022}. This payoff
is included because it tests whether the
joint COS--TT coefficient representation can recover a payoff that cannot be
priced from the one-dimensional arithmetic-basket density.  The purpose is
payoff-coverage validation and literature positioning: it tests a genuinely
joint min-option payoff rather than an arithmetic basket projection.

The bridge uses the correlated GBM convention summarized in
Table~\ref{tab:benchmark-market-specs}. For this homogeneous equicorrelated
case, the reference price can be computed by conditioning on the common
Gaussian factor and evaluating the remaining one-dimensional integral
deterministically. This reference is independent of COS--TT--CHF and remains
available for the extension rows \(d=20\) and \(d=30\).

\begin{table}[htbp]
\centering
\caption{Fast-target COS--TT--CHF validation for the Kastoryano--Pancotti call-on-minimum benchmark.}
\label{tab:kp-min-option-fast-target}
\scriptsize
\setlength{\tabcolsep}{3.2pt}
\begin{tabular}{rrrrcrrr}
\toprule
\(d\) & COS price & Ref. price & Abs. err. &
\shortstack{COS--TT--CHF\\(s)} & \shortstack{KP TF\\(s)} & Rank & CF evals \\
\midrule
2  & 14.86873901 & 14.86874207 & \(3.06\times10^{-6}\) & 0.052 & 0.027 & 9  & \(2{,}250\) \\
3  & 8.97238504  & 8.97242587  & \(4.08\times10^{-5}\) & 0.045 & 0.730 & 9  & \(12{,}950\) \\
4  & 6.15101670  & 6.15101739  & \(6.90\times10^{-7}\) & 0.095 & 6.800 & 13 & \(57{,}024\) \\
5  & 4.53947827  & 4.53947506  & \(3.21\times10^{-6}\) & 0.112 & 10.200 & 13 & \(83{,}520\) \\
6  & 3.51731584  & 3.51731443  & \(1.41\times10^{-6}\) & 0.193 & 13.700 & 17 & \(223{,}440\) \\
7  & 2.82180258  & 2.82180126  & \(1.32\times10^{-6}\) & 0.372 & 52.800 & 18 & \(435{,}792\) \\
8  & 2.32391356  & 2.32391438  & \(8.23\times10^{-7}\) & 0.578 & 63.400 & 17 & \(663{,}840\) \\
9  & 1.95349007  & 1.95349535  & \(5.28\times10^{-6}\) & 0.743 & 74.000 & 17 & \(772{,}320\) \\
10 & 1.66941168  & 1.66941975  & \(8.07\times10^{-6}\) & 0.692 & 84.500 & 17 & \(880{,}800\) \\
15 & 0.89855227  & 0.89855821  & \(5.94\times10^{-6}\) & 1.894 & 326.800 & 21 & \(1{,}897{,}488\) \\
20 & 0.57177577  & 0.57182186  & \(4.61\times10^{-5}\) & 4.266 & -- & 20 & \(3{,}320{,}460\) \\
30 & 0.29803658  & 0.29806560  & \(2.90\times10^{-5}\) & 26.960 & -- & 23 & \(24{,}522{,}048\) \\
\bottomrule
\end{tabular}
\begin{flushleft}
\footnotesize
All rows satisfy an absolute price-error target of \(10^{-4}\) against the
deterministic conditional reference. KP TF denotes the Kastoryano--Pancotti
tensor-Fourier timings reported in their Table I and is included as literature
context, not as a same-machine runtime comparison. Rows \(d=20\) and \(d=30\)
extend beyond the dimensions reported in their Table I.
\end{flushleft}
\end{table}

Table~\ref{tab:kp-min-option-fast-target} shows that COS--TT--CHF reaches the
\(10^{-4}\) pricing regime across all Kastoryano--Pancotti Table I dimensions
and in the additional \(d=20,30\) extension rows. The \(d=2\) row is near the
measurement floor and should not be read as a speed comparison.

\section{High-Dimensional Pricing Results}
We next evaluate COS--TT--CHF on a high-dimensional GBM basket family using an
absolute-deviation target of \(10^{-2}\) against randomized Sobol QMC references.
The benchmark is
the generated Toeplitz GBM arithmetic basket family with equal weights,
\(S_i(0)=100\), \(r=0.02\), \(T=1\), heterogeneous volatilities, and
\(\rho_{ij}=0.7^{|i-j|}\). The corresponding prices, timings, and ranks are reported in
Table~\ref{tab:gbm-fast-pricing}; the complete model and reference protocol are
given in Appendix~\ref{app:table4-fast-gbm-protocol}.

\begin{table}[htbp]
\centering
\caption{High-dimensional GBM arithmetic basket pricing with COS--TT--CHF.}
\label{tab:gbm-fast-pricing}
\scriptsize
\begin{tabular}{rlrrrr}
\toprule
\multicolumn{6}{l}{\textbf{Panel A: \(d=20\) strike grid}}\\
\midrule
\(K\) & Payoff & COS--TT--CHF & QMC ref. & Abs. err. & Rel. err. \\
\midrule
80 & call & 21.694826 & 21.686840 & 0.007986 & \(3.683\times10^{-4}\) \\
80 & put & 0.102775 & 0.102736 & \(3.914\times10^{-5}\) & \(3.810\times10^{-4}\) \\
90 & call & 12.745349 & 12.739923 & 0.005427 & \(4.260\times10^{-4}\) \\
90 & put & 0.958480 & 0.957806 & \(6.744\times10^{-4}\) & \(7.041\times10^{-4}\) \\
100 & call & 5.986084 & 5.982761 & 0.003323 & \(5.555\times10^{-4}\) \\
100 & put & 4.004397 & 4.002631 & 0.001766 & \(4.412\times10^{-4}\) \\
110 & call & 2.192636 & 2.190881 & 0.001755 & \(8.011\times10^{-4}\) \\
110 & put & 10.016131 & 10.012738 & 0.003393 & \(3.389\times10^{-4}\) \\
120 & call & 0.634652 & 0.634116 & \(5.367\times10^{-4}\) & \(8.463\times10^{-4}\) \\
120 & put & 18.263329 & 18.257959 & 0.005369 & \(2.941\times10^{-4}\) \\
\bottomrule
\end{tabular}

\medskip
\begin{tabular}{rlrrrrrrr}
\toprule
\multicolumn{9}{l}{\textbf{Panel B: ATM dimension scaling}}\\
\midrule
\(d\) & Payoff & COS--TT--CHF & QMC ref. & Abs. err. & Total (s) & Tensor (s) & Pricing (s) & Rank \\
\midrule
10 & call & 7.428562 & 7.426033 & 0.002530 & 0.616 & 0.595 & 0.021 & 17 \\
10 & put & 5.447104 & 5.445912 & 0.001192 & 0.616 & 0.595 & 0.021 & 17 \\
15 & call & 6.578794 & 6.575930 & 0.002864 & 0.795 & 0.769 & 0.026 & 13 \\
15 & put & 4.597315 & 4.595765 & 0.001550 & 0.795 & 0.769 & 0.026 & 13 \\
20 & call & 5.986084 & 5.982761 & 0.003323 & 1.841 & 1.807 & 0.035 & 17 \\
20 & put & 4.004397 & 4.002631 & 0.001766 & 1.841 & 1.807 & 0.035 & 17 \\
30 & call & 5.208162 & 5.204752 & 0.003410 & 3.982 & 3.942 & 0.040 & 17 \\
30 & put & 3.226364 & 3.224618 & 0.001746 & 3.982 & 3.942 & 0.040 & 17 \\
\bottomrule
\end{tabular}
\end{table}

\FloatBarrier
Across the diagnostic grid, the maximum absolute deviation from the QMC
reference is \(9.595\times10^{-3}\). The \(d=30\) ATM calculation takes
approximately 3.98 seconds in total, including approximately 0.04 seconds for
pricing after tensor construction.

\subsection{Runtime Scaling Across Models}
We next summarize the runtime scaling of COS--TT--CHF across three terminal
characteristic-function model families: GBM, variance gamma, and normal inverse
Gaussian. For each model and each dimension \(d\in\{5,10,15,20\}\), the method
prices arithmetic-basket calls and puts over strikes
\(K\in\{80,90,100,110,120\}\). The displayed runtimes correspond only to runs
whose maximum absolute deviation from independent randomized Sobol QMC
references is below \(10^{-2}\). The GBM line uses the same generated Toeplitz
market as Table~\ref{tab:gbm-fast-pricing}, but it belongs to a separate run
family with an independent, higher-budget QMC confirmation. The VG and NIG lines use the
specifications in Table~\ref{tab:benchmark-market-specs}. For the VG and NIG
experiments, the out-of-the-money call or put is evaluated directly from the
reconstructed basket density, and its in-the-money counterpart is recovered
from exact basket put--call parity.

\begin{figure}[htbp]
\centering
\includegraphics[width=0.82\textwidth]{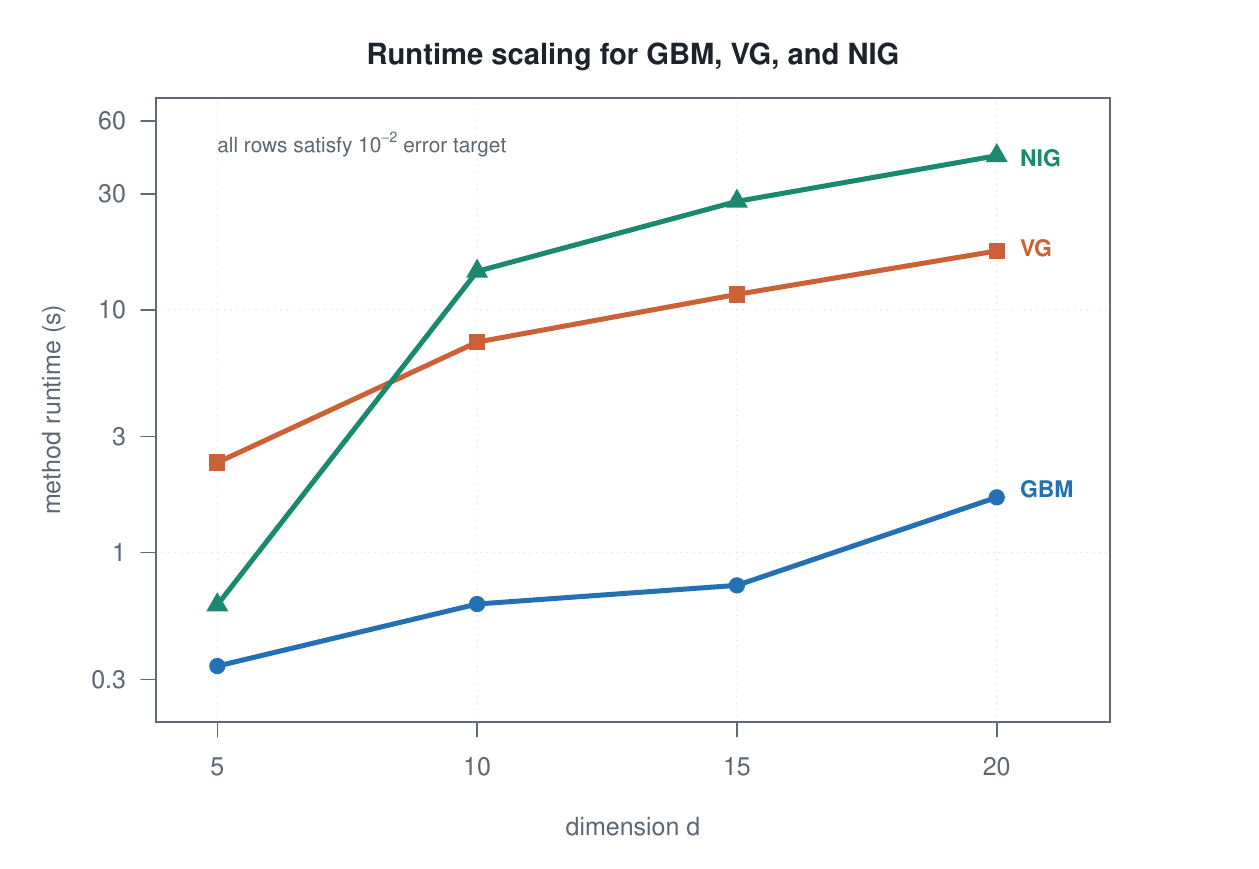}
\caption{Runtime scaling across GBM, VG, and NIG arithmetic-basket pricing problems. Each point reports the COS--TT--CHF method runtime for pricing calls and puts over strikes \(K\in\{80,90,100,110,120\}\). All displayed runs satisfy a maximum absolute pricing-error target of \(10^{-2}\) against independent randomized Sobol QMC references. This conclusion remains unchanged after accounting for the corresponding QMC 95\% confidence half-widths.}
\label{fig:runtime-scaling}
\end{figure}

Figure~\ref{fig:runtime-scaling} separates model difficulty from the accuracy
check reported in the underlying diagnostics. GBM remains inexpensive
throughout the tested range. The heterogeneous VG case has moderate runtime
growth, while the correlated NIG case is the most demanding, with the \(d=20\)
run completing in 43.105 seconds at the same \(10^{-2}\) target. The full
diagnostic table used to build the figure is reported in
Appendix~\ref{app:runtime-scaling-diagnostics}.

\section{Common-Heston Runtime Stress Test}
As a stochastic-volatility stress test, we also consider a common-factor affine
Heston model. Unlike the GBM QMC reference in
Table~\ref{tab:gbm-fast-pricing}, the QMC reference used here simulates paths
and assesses time-step sensitivity, while COS--TT--CHF uses the exact affine
terminal joint characteristic function.
This experiment is therefore designed to stress a regime where terminal
characteristic functions are available but path simulation remains expensive:
QMC must resolve sampling error and assess Euler discretization sensitivity,
whereas COS--TT--CHF works directly with the terminal transform.

\begin{table}[htbp]
\centering
\caption{Common-Heston runtime stress test against coupled randomized Sobol QMC.}
\label{tab:heston-qmc-stress}
\scriptsize
\setlength{\tabcolsep}{2.5pt}
\begin{tabular}{rrrrrrrrr}
\toprule
\(d\) & COS (s) & Max rank & Mass err. & COS err & QMC (s) & QMC half & Max step-diff. diag. & QMC/COS \\
\midrule
2 & 0.068 & 13 & \(2.3\times10^{-5}\) & 0.002072 & 58.576 & 0.004158 & \(3.011\times10^{-4}\) & 865.31 \\
5 & 4.690 & 22 & \(5.6\times10^{-5}\) & 0.005456 & 102.838 & 0.003397 & \(2.945\times10^{-4}\) & 21.93 \\
10 & 17.457 & 24 & \(9.5\times10^{-5}\) & 0.003318 & 219.373 & 0.002445 & \(2.605\times10^{-4}\) & 12.57 \\
20 & 62.613 & 28 & \(6.7\times10^{-5}\) & 0.002630 & 501.728 & 0.002401 & \(2.740\times10^{-4}\) & 8.01 \\
\bottomrule
\end{tabular}
\begin{flushleft}
\footnotesize
The target is \(10^{-2}\) over the full call/put strike grid, and all rows
satisfy the QMC qualification rule defined in
Appendix~\ref{app:heston-qmc-protocol}. QMC half and Max step-diff. diag. report
the maximum sampling and coupled-refinement diagnostics, respectively. Mass
err. is the recovered basket-density mass error, COS err is the maximum
absolute difference from the QMC means, and QMC/COS is the runtime ratio.
\end{flushleft}
\end{table}

The complete model and QMC path-simulation protocol are specified in
Appendix~\ref{app:heston-qmc-protocol}. All rows pass the stated QMC
qualification rule. COS--TT--CHF remains within the grid-level \(10^{-2}\)
target in all dimensions, and the mass errors are below \(10^{-4}\).

The runtime ratios range from about \(865\) at \(d=2\) to \(8\) at \(d=20\).
The advantage is largest at low dimension, where COS--TT--CHF has little
tensor-construction overhead while QMC still pays the full path-simulation and
coupled-refinement cost. As \(d\) increases, the maximum TT rank rises from
\(13\) to \(28\), and the COS--TT--CHF construction becomes more expensive.
Nevertheless, it remains faster than the QMC reference throughout the reported
grid.

This is a scoped model-regime result, not a claim of universal dominance over
QMC. It shows that, when an affine terminal characteristic function is
available but QMC must simulate stochastic-volatility paths, COS--TT--CHF can
turn the transform representation into a substantially faster grid-level
pricing calculation. The same mechanism can arise in other affine
stochastic-volatility settings, such as multi-factor Heston or Bates-type
jump-diffusion models, where path-based QMC must simulate latent factors and
assess time-step sensitivity. Terminally sampled GBM baskets remain a different
regime.

\section{Multi-Strike Surface Reuse}
\label{sec:multi-strike-reuse}
This experiment measures how efficiently the compressed representation can be
reused across a full strike grid after the one-time TT construction. It uses the reported
\(d=20\) heterogeneous VG specification from
Table~\ref{tab:benchmark-market-specs}. The purpose is not an end-to-end
comparison against QMC. QMC is used only to verify the accuracy of the computed
surface. The corresponding strike-reuse timings and QMC deviations are
reported in Table~\ref{tab:vg-multistrike-reuse}.

The practical configuration uses \(W=28\), \(M=56\), COS order \(N=64\),
and TT rank cap \(22\). The out-of-the-money call or put is evaluated directly
from the reconstructed density, and its in-the-money counterpart is recovered
from exact basket put--call parity.

The reuse mechanism is the prefix representation
\eqref{eq:basket-call-prefix}--\eqref{eq:basket-put-prefix}. Once the
compressed tensor has produced the basket density \(f_H\), the cumulative
quantities \(L_0\), \(L_1\), \(T_0\), and \(T_1\) are fixed for the whole
strike surface. A new strike \(K\) then changes only the scalar evaluations
\(L_0(K)\) and \(L_1(K)\) in the call and put formulas, not the TT-cross
construction, the CHF-to-COS transform, or the density reconstruction.

Accordingly, Table~\ref{tab:vg-multistrike-reuse} focuses on marginal lookup
cost, amortized per-strike cost, and QMC deviation. The fixed setup,
density-reconstruction, rank, and lookup range are stated in the table note.
The amortized cost falls from 19.40 seconds for a single strike to 0.24
seconds per strike for the full 81-strike surface. Over this surface, the
maximum parity-consistent deviation from the randomized Sobol QMC means is
\(7.652\times10^{-3}\), and the maximum QMC 95\% confidence half-width is
\(8.479\times10^{-4}\). Their conservative sum is
\(8.500\times10^{-3}<10^{-2}\), so the practical configuration passes the
one-cent target after accounting for QMC sampling uncertainty.

\begin{table}[htbp]
\centering
\caption{Strike-reuse timing for the heterogeneous VG basket case.}
\label{tab:vg-multistrike-reuse}
\scriptsize
\begin{tabular}{rrrr}
\toprule
Strikes & Marginal/strike (s) & Amort./strike (s) & QMC dev. \\
\midrule
1 & 0 & 19.395 & \(5.333\times10^{-3}\) \\
5 & \(2.948\times10^{-6}\) & 3.879 & \(5.549\times10^{-3}\) \\
11 & \(1.271\times10^{-7}\) & 1.763 & \(7.521\times10^{-3}\) \\
21 & \(9.580\times10^{-8}\) & 0.924 & \(7.521\times10^{-3}\) \\
41 & \(4.531\times10^{-8}\) & 0.473 & \(7.521\times10^{-3}\) \\
81 & \(4.426\times10^{-8}\) & 0.239 & \(7.652\times10^{-3}\) \\
\bottomrule
\end{tabular}
\begin{flushleft}
\footnotesize
The one-time setup cost is 19.386 seconds and the basket-density
reconstruction time is 0.009278 seconds.
Marginal/strike is the additional lookup time per extra strike relative to the
one-strike row. QMC dev. is the maximum parity-consistent call/put deviation
against randomized Sobol QMC means. The QMC reference uses 64 independently
scrambled Sobol point sets with \(2^{17}\) points per scramble, for
\(2^{23}\) points in total. The same terminal samples are reused
across all strikes and both payoffs, and the 95\% confidence intervals are
computed from variation across scrambles.
\end{flushleft}
\end{table}

\section{Discussion}

The experiments support COS--TT--CHF as an efficient high-dimensional solver in
the tested regimes, built directly from the joint characteristic function. The
method is useful when the sampled characteristic-function tensor has moderate
effective TT ranks on a resolved frequency grid. This condition is not automatic: an exact
TT representation exists for any finite tensor if the ranks are allowed to grow,
but the computational advantage comes only when the ranks remain small enough
for TT-cross sampling, coefficient conversion, and payoff contraction to be
cheaper than the corresponding full-grid calculation. This point is central to
the comparison with Schaap's original COS--TT--CHF experiments, where direct
compression of sampled characteristic-function tensors was difficult because
smooth tensors may still be highly oscillatory. The present workflow addresses
this difficulty through explicit numerical controls. Rank, held-out tensor
checks, mass and parity diagnostics, and sensitivity sweeps are therefore part of
the evidence, not after-the-fact checks.

The runtime comparisons should be read in the same representation-building
setting. QMC is a strong benchmark because it is flexible, often fast, and gives
independent reference values with confidence intervals. Direct COS remains
competitive in low dimensions, where the full tensor grid is still small. For
isolated low-dimensional prices, direct COS or QMC may be preferable because the
TT setup cost has little opportunity to amortize. COS--TT--CHF addresses a
different numerical object: it constructs a compressed spectral representation
from the joint characteristic function. In the local direct-COS comparison, the
crossover appears between \(d=2\) and \(d=4\). In the min-option bridge and
common-Heston stress test, the reported timings favor COS--TT--CHF in the
tested dimensions beyond \(d=2\), subject to the comparison conventions stated
in the tables.

The payoff and Greek calculations illustrate what this representation can
support after it has been built. For arithmetic baskets, the joint COS
coefficient TT is projected to a one-dimensional basket density before strike
evaluation. For min/max options, the payoff cannot be reduced in the same way.
the calculation uses rectangle probabilities and tail integrals from the joint
coefficient tensor. For GBM baskets, component Delta and component Vega are
obtained by differentiating the characteristic function and propagating the
derivative tensors through the same transform pipeline. These tests support the
interpretation of COS--TT--CHF as a reusable spectral representation, but they
do not yet constitute a systematic study of Greek surfaces, higher-order Greeks,
or the accuracy of sensitivities across model classes.

The results also point to several extensions. The paper gives numerical
evidence, not a general rank theory for characteristic-function tensors. The
frequency grid and truncation interval remain important numerical choices. The
benchmark families cover several useful regimes, but they do not exhaust the
behavior of affine, Lévy, or stochastic-volatility models. Natural extensions
include a broader rank and grid study, additional Greeks and hedging quantities,
partial reuse across related initial-spot scenarios, and early-exercise pricing
for Bermudan and American-style contracts. The last direction is natural because
early-exercise pricing repeatedly evaluates continuation values, which is close
in spirit to reusing a spectral representation across payoff evaluations and
exercise dates. Fourier time-stepping methods have already addressed
early-exercise contracts under Lévy models, and COS methods were subsequently
extended to early-exercise, barrier, Bermudan, and path-dependent settings
\citep{lord2008fft,fangoosterlee2009earlyexercise,zhangoosterlee2012bermudan,
zhangoosterlee2014asian}. Related spectral Padé and convolution methods provide
further evidence that transform-based continuation-value evaluation is a viable
route for early-exercise pricing \citep{chan2018,chan2020,chanhale2020}.
Finally, recent work on TT methods for
multidimensional inverse Laplace transforms \citep{mikkelsen2026laplace}
suggests a possible complementary route for models or payoffs more naturally
represented through Laplace transforms than through Fourier transforms.

\section{Conclusion}

This paper develops COS--TT--CHF into a numerical workflow for European
multi-asset pricing under characteristic-function models. The workflow samples
the joint characteristic function on a tensor-product frequency grid, compresses
the sampled tensor by TT-cross, converts the compressed object into COS
coefficient cores, and evaluates basket, min/max, multi-strike, and selected
Greek quantities from the resulting representation.

The experiments reproduce published basket benchmarks, validate the
Kastoryano--Pancotti min-option bridge within the stated tolerances, and extend
the tests to \(d=30\) for GBM and \(d=20\) for VG, NIG, and common-Heston
benchmark families. The same compressed representation supports fast post-setup
strike-grid evaluation and selected component Delta and component Vega
calculations.

The experiments show that, in the tested regimes, the compressed
characteristic-function representation can be built, checked, and reused for
multi-asset pricing tasks. The diagnostics reported throughout the paper are
essential to this conclusion, because they make the numerical regime visible.
They identify whether the frequency grid, TT ranks, coefficient conversion, and
payoff contractions are resolving the representation used for pricing and
selected sensitivities. These results position COS--TT--CHF as a
high-dimensional characteristic-function solver in its own right. It builds a
tensor-compressed COS representation from the joint characteristic function,
checks that representation numerically, and reuses it for prices, strike
surfaces, payoff-specific contractions, diagnostics, and selected sensitivities.

\section*{Data and Code Availability}

The implementation code is proprietary and is not publicly released.  The paper
reports the mathematical formulation, payoff definitions, algorithmic workflow,
numerical parameters, reference-pricing procedures, and diagnostics used to
interpret the reported pricing, runtime, and Greek results.

\section*{Funding}
This research was funded by the Carlsberg Foundation (grant no. CF22-1030) and
the Novo Nordisk Foundation Quantum Computing Programme (grant no.
NNF22SA0081175).

\section*{Disclosure Statement}

The authors report no conflict of interest.

\appendix

\section{Runtime Scaling Diagnostics}

\label{app:runtime-scaling-diagnostics}
Table~\ref{tab:runtime-scaling-diagnostics} reports the diagnostics used to
construct Figure~\ref{fig:runtime-scaling}. Each row summarizes the maximum
absolute error
over calls and puts with strikes
\(K\in\{80,90,100,110,120\}\). The independent randomized Sobol QMC protocol
provides the reference prices. The figure reports the COS--TT--CHF method
runtimes. The ``Worst row'' column identifies the strike/payoff pair attaining
the maximum error, so each runtime point can be read together with its accuracy
bottleneck.

For all three model families and all reported dimensions, the QMC reference
uses 32 independently scrambled Sobol point sets with
\(2^{17}\) points per scramble, for \(2^{22}\)
points per model--dimension case. This is a separate uniform high-precision
QMC confirmation of the cached runtime-scaling runs, not the \(2^{16}\)-point
reference experiment used for Table~\ref{tab:gbm-fast-pricing}. Consequently,
the GBM errors and method timings in the two tables need not coincide. The
reported confidence intervals
are 95\% \(t\)-intervals computed from the variation across scrambles. The
maximum absolute errors in Table~\ref{tab:runtime-scaling-diagnostics} are
computed relative to the QMC means. The conclusion is unchanged after
accounting for QMC uncertainty: adding the corresponding 95\% confidence
half-width to each absolute deviation leaves all 120 reported strike/payoff
rows below \(10^{-2}\), with a maximum value of \(0.008694\).

\begin{table}[htbp]
\centering
\caption{Underlying diagnostics for the runtime-scaling figure. All rows satisfy the \(10^{-2}\) maximum absolute error target against independent randomized Sobol QMC references.}
\label{tab:runtime-scaling-diagnostics}
\scriptsize
\setlength{\tabcolsep}{3.5pt}
\begin{tabular}{lrrrrl}
\toprule
Model & \(d\) & Max err. & Method (s) & Rank & Worst row \\
\midrule
GBM & 5 & 0.005553 & 0.340 & 13 & call \(K=80\) \\
GBM & 10 & 0.006078 & 0.613 & 17 & call \(K=80\) \\
GBM & 15 & 0.006602 & 0.732 & 13 & call \(K=80\) \\
GBM & 20 & 0.008335 & 1.687 & 17 & call \(K=80\) \\
\midrule
VG & 5 & 0.004416 & 2.347 & 22 & call \(K=110\) \\
VG & 10 & 0.005157 & 7.357 & 22 & call \(K=110\) \\
VG & 15 & 0.004859 & 11.568 & 22 & put \(K=110\) \\
VG & 20 & 0.005787 & 17.473 & 22 & put \(K=100\) \\
\midrule
NIG & 5 & 0.003446 & 0.608 & 25 & call \(K=110\) \\
NIG & 10 & 0.001092 & 14.394 & 32 & put \(K=100\) \\
NIG & 15 & 0.002958 & 27.952 & 32 & call \(K=100\) \\
NIG & 20 & 0.001210 & 43.105 & 32 & call \(K=110\) \\
\bottomrule
\end{tabular}
\end{table}
\FloatBarrier

For the NIG rows, the log-price truncation multiplier is \(L_1=6.5\) for
\(d=5,10,15\) and \(L_1=8\) for \(d=20\).

\section{Benchmark Conventions and Source Artifacts}
\label{app:benchmark-conventions}
Table~\ref{tab:medium-dimensional-validation} is assembled from the Bayer et al. and
Junike--Stier bridge artifacts used to generate the reported validation rows.
All rows are European arithmetic basket-put benchmark cases, but the two
literature sources use different conventions and should not be merged into a
single parameter description.

\begin{table}[htbp]
\centering
\caption{Medium-dimensional benchmark conventions for the Bayer et al. and Junike--Stier validation rows.}
\label{tab:medium-dimensional-benchmark-conventions}
\scriptsize
\begin{tabular}{@{}>{\raggedright\arraybackslash}p{0.18\textwidth}
                  >{\raggedright\arraybackslash}p{0.36\textwidth}
                  >{\raggedright\arraybackslash}p{0.36\textwidth}@{}}
\toprule
Quantity & Bayer et al. rows & Junike--Stier rows \\
\midrule
Payoff & European arithmetic average basket put & Arithmetic basket put \\
Source tables & Bayer et al. Tables 4.1, 4.2, and 4.6 & Junike--Stier Tables 4 and 5 \\
Models & GBM and VG & BS/GBM and VG \\
Dimensions & \(d=4,6\) & \(d=2,4\) \\
Spots & \(S_i(0)=100\) & \(S_i(0)=100/d\) for the unweighted basket convention \\
Maturity/rate & \(T=1\), \(r=0\) & \(T=1\), \(r=0\) \\
Correlation & Independent assets, \(C=I_d\) & BS/GBM equicorrelation \(0.5\) and VG identity correlation \\
Basket weights & Equal weights \(1/d\) & Unweighted sum convention in the bridge implementation \\
Strikes & \(K=100\) for \(d=4\) and \(K=60\) for \(d=6\) & \(K=100\) \\
Reference comparison & Published Bayer et al. reference prices and OD+ASGQ relative errors from their Table~4.6 & Published references with absolute-error target about \(10^{-2}\) \\
Local method & COS--TT--CHF paper profile, three repeats & COS--TT--CHF on the same cases \\
Baseline runtime & OD+ASGQ (reported), contextual only & COS-iv (local tuned), tuned to first tested configuration hitting the target \\
\bottomrule
\end{tabular}
\end{table}
\clearpage

\section{High-Dimensional GBM Protocol}
\label{app:table4-fast-gbm-protocol}
Table~\ref{tab:gbm-fast-pricing} uses the reported \(10^{-2}\)-target
COS--TT--CHF settings against independent physical-domain randomized Sobol QMC
references.
The generated model is the Toeplitz GBM basket case described below.
Table~\ref{tab:table4-fast-gbm-market-parameters} fixes the market and payoff
parameters for the high-dimensional GBM experiment, with
\(x_i=(i-1)/(d-1)\).

\begin{table}[htbp]
\centering
\caption{GBM market and payoff parameters for the fast high-dimensional pricing table.}
\label{tab:table4-fast-gbm-market-parameters}
\small
\begin{tabular}{lp{0.66\textwidth}}
\toprule
Quantity & Value \\
\midrule
Model family & GBM \\
Dimensions & \(d=10,15,20,30\) \\
Spot & \(S_i(0)=100\) \\
Rate and maturity & \(r=0.02\), \(T=1\) \\
Basket weights & Equal weights \\
Payoffs & Arithmetic basket call and put \\
Strike grid & \(K=80,90,100,110,120\) \\
ATM strike & \(K=100\) \\
Volatility rule & \(\sigma_i=0.18+(0.30-0.18)x_i+0.015\sin(\pi x_i)\) \\
Correlation rule & \(\rho_{ij}=0.7^{|i-j|}\) \\
\bottomrule
\end{tabular}
\end{table}

Table~\ref{tab:table4-fast-qmc-protocol} records the independent randomized
Sobol reference protocol used to check the prices in
Table~\ref{tab:gbm-fast-pricing}.

\begin{table}[htbp]
\centering
\caption{Randomized Sobol QMC reference protocol for the fast high-dimensional GBM table.}
\label{tab:table4-fast-qmc-protocol}
\small
\begin{tabular}{lp{0.66\textwidth}}
\toprule
Quantity & Value \\
\midrule
QMC method
& Randomized Sobol simulation of the exact terminal GBM distribution with
LMS+digital-shift scrambling \citep{sobol1967,matousek1998} \\
Marginal transform
& Componentwise inverse standard-normal CDF \\
Covariance factorization
& PCA with eigenvalues ordered from largest to smallest \\
Scrambles
& 32 independently scrambled Sobol sequences \\
Paths per scramble
& \(2^{16}\) \\
Total paths
& \(2^{21}\) per dimension; the same terminal samples are reused across
strikes and payoffs \\
Confidence intervals
& 95\% \(t\)-intervals constructed from variation across the independent
scrambles; these intervals are reported as diagnostics but are not used as
the declared validation target \\
\bottomrule
\end{tabular}
\end{table}

The resulting COS--TT--CHF diagnostics are reported in
Table~\ref{tab:table4-fast-gbm-diagnostics}.

\begin{table}[htbp]
\centering
\caption{Diagnostics for the fast high-dimensional GBM table.}
\label{tab:table4-fast-gbm-diagnostics}
\scriptsize
\begin{tabular}{rrrrrrrr}
\toprule
\(d\) & Mass & Mean & Max abs. err. & Tol. hit & Monotone & Max parity & TT val eps \\
\midrule
10 & 1.00025212 & 102.046699 & 0.006226 & yes & yes & 0.006 & \(1.313\times10^{-7}\) \\
15 & 1.00028147 & 102.049655 & 0.006822 & yes & yes & 0.007 & \(5.477\times10^{-6}\) \\
20 & 1.00032595 & 102.054315 & 0.007986 & yes & yes & 0.008 & \(5.175\times10^{-7}\) \\
30 & 1.00040239 & 102.062071 & 0.009595 & yes & yes & 0.010 & \(3.084\times10^{-6}\) \\
\bottomrule
\end{tabular}
\begin{flushleft}
\footnotesize
The sampled TT entry errors are \(1.090\times10^{-6}\), \(7.136\times10^{-6}\),
\(1.532\times10^{-6}\), and \(6.567\times10^{-7}\) for
\(d=10,15,20,30\), respectively. ``Mass'' and ``Mean'' are the mass and first
moment of the reconstructed basket density. ``Max abs. err.'' and ``Max
parity'' are maxima over the five-strike call/put grid. ``Tol. hit'' means that
the maximum absolute price error is at most \(10^{-2}\); ``Monotone'' checks
that calls are nonincreasing and puts are nondecreasing with strike. ``TT val
eps'' is the final held-out validation residual returned by TT-cross.
\end{flushleft}
\end{table}
\FloatBarrier

\section{Common-Heston Stress-Test Protocol}
\label{app:heston-qmc-protocol}
This section specifies the common-factor affine Heston model and the coupled
QMC protocol underlying Table~\ref{tab:heston-qmc-stress}.
Table~\ref{tab:heston-qmc-market-parameters} lists the market and model
parameters, with \(x_i=(i-1)/(d-1)\).

\begin{table}[htbp]
\centering
\caption{Common-Heston market and model parameters for the QMC stress test.}
\label{tab:heston-qmc-market-parameters}
\small
\begin{tabular}{lp{0.66\textwidth}}
\toprule
Quantity & Value \\
\midrule
Payoffs & Arithmetic basket calls and puts \\
Dimensions & \(d=2,5,10,20\) \\
Strikes & \(K=80,90,100,110,120\) \\
Spot and weights & \(S_i(0)=100\) with equal weights \(w_i=1/d\) \\
Rate and maturity & \(r=0.02\), \(T=1\) \\
Variance parameters & \(\kappa=2\), \(\theta=0.04\), \(v_0=0.04\), \(\xi=0.07\) \\
Common leverage & \(\beta_i=-0.25\) for all assets \\
Asset loading curve & \(a_i=(0.18+0.12x_i+0.015\sin(\pi x_i))/\sqrt{0.04}\) \\
Initial effective volatilities & \(a_i\sqrt{v_0}\) range from 18\% to 30\% \\
Residual correlation & Toeplitz decay \(0.45^{|i-j|}\), scaled by the idiosyncratic variances so that \(\operatorname{Cov}(dW)=\operatorname{Cov}(dR)+\beta\beta^\top dt\) \\
\bottomrule
\end{tabular}
\end{table}
\FloatBarrier

The simulated log-price and variance dynamics are
\begin{align*}
  dX_i(t)
  &=
  \left(r-\frac{1}{2}a_i^2 V_t\right)dt
  + a_i\sqrt{V_t}\,dW_i(t), \\
  dV_t
  &=
  \kappa(\theta-V_t)dt+\xi\sqrt{V_t}\,dB_t, \\
  dW_i(t)
  &=
  \beta_i\,dB_t + dR_i(t),
\end{align*}
where \(R_i\) denotes the residual asset-shock component.
For \(R=(R_1,\ldots,R_d)\), the residual covariance is \(\operatorname{Cov}(dR_i,dR_j)=\sqrt{1-\beta_i^2}\,0.45^{|i-j|}\sqrt{1-\beta_j^2}\,dt\); thus, each component of \(W=(W_1,\ldots,W_d)\) has unit variance, and its instantaneous correlation matrix is \(\operatorname{Cov}(dR)/dt+\beta\beta^\top\).

Table~\ref{tab:heston-qmc-protocol} records the QMC path-simulation protocol
and the pass rule used for the Common-Heston runtime comparison.

\begin{table}[htbp]
\centering
\caption{QMC protocol for the common-Heston runtime stress test.}
\label{tab:heston-qmc-protocol}
\small
\begin{tabular}{lp{0.66\textwidth}}
\toprule
Quantity & Value \\
\midrule
QMC method & Randomized Sobol path simulation with LMS+digital-shift scrambling \citep{sobol1967,matousek1998} \\
Scrambles & 32 independent scrambles \\
Paths per scramble & \(2^{16}\) \\
Total terminal paths & \(2^{21}\) \\
Simulation scheme & Log-Euler asset update with full-truncation Euler variance discretization \citep{lord2010biased} \\
Fine/coarse steps & 128 fine steps and 64 coarse steps \\
Step-difference diagnostic & For \(D=\widehat P_{128}-\widehat P_{64}\), \(|\bar D|+h_{0.95}(D)\), where \(\widehat P_n\) is the option-price estimate from an \(n\)-step simulation and \(h_{0.95}(D)\) is the 95\% confidence half-width across randomized QMC scrambles. Coarse paths are coupled to fine paths by aggregating the same Brownian increments \\
QMC target & \(10^{-2}\) over the full call/put strike grid \\
Target pass rule & Maximum QMC 95\% half-width and maximum coupled step-difference diagnostic over the strike/payoff grid both below \(5\times10^{-3}\) \\
\bottomrule
\end{tabular}
\end{table}
\FloatBarrier

The coupled step-difference diagnostic is a refinement check, not a rigorous
upper bound on the remaining discretization bias of the 128-step estimator.

\section{Additional Sensitivity Sweeps}
\label{app:additional-stability-sweeps}
This appendix reports the Gaussian counterpart to the main-text VG sensitivity
check. As in Table~\ref{tab:vg-numerical-stability}, one numerical control is
varied at a time while the remaining settings are held fixed. The diagnostics
have the same definitions: ``Ref. dev.'' is the maximum absolute call/put
deviation from a higher-resolution internal reference run, ``Baseline change'' is the
maximum change from the default row, and the final columns report density mass
and put--call parity checks. All prices use \(d=20\) and \(K=100\). The
baseline uses rank cap 20, frequency half-width \(W=24\), \(M=48\) frequency
nodes per axis, COS order \(N=64\), and TT-cross seed 42. The GBM sweep is
placed in the appendix because it confirms the same sensitivity pattern as the
main VG sweep in a simpler Gaussian setting.

\begin{table}[htbp]
\centering
\caption{Sensitivity of GBM \(d=20\) COS--TT--CHF prices to one-at-a-time perturbations of numerical controls. This appendix table provides the Gaussian baseline corresponding to the main-text VG sensitivity sweep.}
\label{tab:gbm-numerical-stability}
\scriptsize
\begin{tabular}{llrrrrrr}
\toprule
Control & Setting & Call & Put & Ref. dev. & Baseline change & Mass err. & Parity resid. \\
\midrule
baseline & default & 5.976856 & 3.999636 & \(1.675\times10^{-3}\) & \(0.000\times10^{0}\) & \(8.051\times10^{-4}\) & \(2.914\times10^{-3}\) \\
rank cap & 16 & 5.976834 & 3.999627 & \(1.697\times10^{-3}\) & \(2.196\times10^{-5}\) & \(8.057\times10^{-4}\) & \(2.926\times10^{-3}\) \\
rank cap & 24 & 5.976856 & 3.999637 & \(1.674\times10^{-3}\) & \(2.608\times10^{-7}\) & \(8.051\times10^{-4}\) & \(2.913\times10^{-3}\) \\
frequency width & \(W=20\) & 5.977532 & 3.999938 & \(9.987\times10^{-4}\) & \(6.759\times10^{-4}\) & \(7.014\times10^{-4}\) & \(2.539\times10^{-3}\) \\
frequency width & \(W=28\) & 5.975910 & 3.999166 & \(2.620\times10^{-3}\) & \(9.456\times10^{-4}\) & \(9.351\times10^{-4}\) & \(3.389\times10^{-3}\) \\
nodes/axis & \(M=40\) & 5.971926 & 3.997390 & \(6.604\times10^{-3}\) & \(4.930\times10^{-3}\) & \(1.494\times10^{-3}\) & \(5.597\times10^{-3}\) \\
nodes/axis & \(M=56\) & 5.979069 & 4.000715 & \(5.384\times10^{-4}\) & \(2.213\times10^{-3}\) & \(4.938\times10^{-4}\) & \(1.779\times10^{-3}\) \\
COS order & \(N=48\) & 5.977011 & 3.999667 & \(1.519\times10^{-3}\) & \(1.558\times10^{-4}\) & \(8.036\times10^{-4}\) & \(2.788\times10^{-3}\) \\
COS order & \(N=96\) & 5.976855 & 3.999636 & \(1.675\times10^{-3}\) & \(2.274\times10^{-7}\) & \(8.051\times10^{-4}\) & \(2.914\times10^{-3}\) \\
TT-cross seed & 1729 & 5.976856 & 3.999637 & \(1.675\times10^{-3}\) & \(3.811\times10^{-8}\) & \(8.051\times10^{-4}\) & \(2.914\times10^{-3}\) \\
TT-cross seed & 1910 & 5.976856 & 3.999636 & \(1.675\times10^{-3}\) & \(9.548\times10^{-8}\) & \(8.051\times10^{-4}\) & \(2.913\times10^{-3}\) \\
\bottomrule
\end{tabular}
\end{table}
\clearpage

The GBM sweep shows the same qualitative pattern as the VG case: changes in
rank cap and TT-cross seed have negligible effect around the baseline, while
frequency-window and node-count choices drive the larger deviations. This
supports the interpretation that the main residual variation in these sensitivity
checks is Fourier/COS discretization rather than TT-cross seed variability.

\FloatBarrier

\end{document}